\documentclass[PAPER, atlasdraft=false, coverpage=false,cernpreprint=true, USenglish, texmf, orcidlogo]{atlasdoc}
 
\usepackage[subcaption]{atlaspackage}
\usepackage{atlasbiblatex}
 
\usepackage{atlasphysics}

\graphicspath{{logos/}{figures/}}
 
\usepackage{ANA-FTAG-2023-02-PAPER-defs}

\AtlasTitle{Calibration of a soft secondary vertex tagger using proton--proton collisions at $\sqrt{s}=$ 13 TeV with the ATLAS detector}

\AtlasAbstract{
Several processes studied by the ATLAS experiment at the Large Hadron Collider produce low-momentum $b$-flavored hadrons in the final state.
This paper describes the calibration of a dedicated tagging algorithm that identifies $b$-flavored hadrons outside of hadronic jets by reconstructing the soft secondary vertices originating from their decays.
The calibration is based on a proton--proton collision dataset at a center-of-mass energy of 13 TeV corresponding to an integrated luminosity of 140 fb$^{-1}$. Scale factors used to correct the algorithm's performance in simulated events are extracted for the $b$-tagging efficiency and the mistag rate of the algorithm using a data sample enriched in $t\bar{t}$ events.
Several orthogonal measurement regions are defined, binned as a function of the multiplicities of soft secondary vertices and jets containing a $b$-flavored hadron in the event.
The mistag rate scale factors are estimated separately for events with low and high average number of interactions per bunch crossing.
The results, which are derived from events with low missing transverse momentum, are successfully validated in a phase space characterized by high missing transverse momentum
and therefore are applicable to new physics searches carried out in either phase space regimes.
}

\AtlasRefCode{FTAG-2023-02}
 
\PreprintIdNumber{CERN-EP-2024-108}

\AtlasJournalRef{\PRD 110 (2024) 032015}
\AtlasDOI{10.1103/PhysRevD.110.032015}

\hypersetup{pdftitle={ATLAS document},pdfauthor={The ATLAS Collaboration}}
 
\begin{document}
 
\maketitle

\section{Introduction}

The accurate reconstruction and identification of particles traversing the ATLAS detector is a fundamental aspect of the experiment's successful physics program. This includes the identification of jets that contain a $b$-flavored hadron ($B$-hadron in the following), referred to as \bjets, which is crucial for enabling precise measurements and discoveries of processes involving heavy particles that decay into $b$-quarks. The distinctive decay patterns and long lifetimes of $B$-hadrons, as well as the fragmentation properties of $b$-quarks, provide a powerful handle for distinguishing \bjets from jets originating from lighter quarks or gluons.
The ATLAS Collaboration has developed various algorithms to identify \bjets, referred to as \btagging\ algorithms~\cite{PERF-2012-04,FTAG-2018-01,FTAG-2019-07}, which are essential for analyzing the data recorded at the Large Hadron Collider (LHC).
 
The use of jets as a starting point for $b$-tagging imposes constraints on the energy of both the $B$-hadron and the surrounding hadronic activity. Standard $b$-tagging techniques in ATLAS are applicable to jets with transverse momenta (\pT) of at least 20~\GeV.
However, many scenarios beyond the Standard Model (SM) predict final states with $B$-hadrons at lower \pT, such that their identification would greatly improve the associated analysis sensitivity.

One class of processes that could give rise to low-\pT $B$-hadrons is the production of the supersymmetric (SUSY) partners of third generation quarks (top and bottom squarks). In particular, compressed SUSY scenarios~\cite{Golfand:1971iw,Volkov:1973ix,Wess:1974tw,Wess:1974jb,Ferrara:1974pu,Salam:1974ig}, where either the top or the bottom squark is nearly mass-degenerate with the lightest
SUSY particle, are favored in scenarios of electroweak baryogenesis.
The low \pt of the squark decay products makes such scenarios difficult to separate from background. The \pt\ spectrum of the final state particles in such events may peak as low as $4\;\GeV$ depending on the mass compression.
 
Other examples of searches for physics beyond the SM that may benefit from low-\pT $B$-hadron identification are searches for exotic decays of the Higgs boson~\cite{HDBS-2018-47} into new light pseudoscalar particles. Such processes can involve the production of up to six $b$-quarks in the final state~\cite{Robens_2020}.
Depending on the relative mass difference between the new pseudoscalar states and the Higgs boson, the
\pt spectrum of the lowest-\pT $b$-quarks can peak as low as $10\;\GeV$. The ability to identify such low-\pt $b$-quarks is crucial for the analysis performance in discriminating the signal from the SM background.

The ATLAS Collaboration has developed algorithms
tagging low-\pT $B$-hadrons arising in compressed bottom squark~\cite{SUSY-2018-34}
and top squark~\cite{SUSY-2018-07} searches, with $b$-quarks in the final state. The CMS Collaboration has also successfully implemented similar techniques in the context
of compressed-SUSY scenarios~\cite{CMS-SUS-16-049}.
This paper focuses on the calibration of the Track-Cluster-based Low-\pT
Vertex Tagger (TC-LVT). This algorithm is  based on the reconstruction of secondary vertices outside of jets identified by traditional clustering algorithms
and targets the reconstruction of $B$-hadrons with \pT between $5$ and $20\;\GeV$.
 
A data sample containing predominantly top-antitop-quark (\ttbar) events is selected, where both top quarks decay into a $W$ boson and a $b$-quark, followed by
a subsequent decay of each $W$ boson into either an electron or a muon and a neutrino.
 
This paper is organized as follows. In Section~\ref{sec:detector} the ATLAS detector is described.
The data and simulated samples used in the calibration are described in Section~\ref{sec:datasetMC}.
Section~\ref{sec:objects} summarizes the reconstruction of tracks, electrons, muons and jets, followed by a
description of the TC-LVT algorithm in Section~\ref{sec:tclvtdefperf}.
Section~\ref{sec:analysis} describes the event selection and calibration method, systematic uncertainties and results. The conclusions are given in  Section~\ref{sec:conclusion}.

\section{The ATLAS detector}
\label{sec:detector}
 
\newcommand{\AtlasCoordFootnote}{
ATLAS uses a right-handed coordinate system with its origin at the nominal interaction point (IP)
in the center of the detector and the \(z\)-axis along the beam pipe.
The \(x\)-axis points from the IP to the center of the LHC ring,
and the \(y\)-axis points upwards.
Polar coordinates \((r,\phi)\) are used in the transverse plane,
\(\phi\) being the azimuthal angle around the \(z\)-axis.
The pseudorapidity is defined in terms of the polar angle \(\theta\) as \(\eta = -\ln \tan(\theta/2)\).
Angular distance is measured in units of \(\Delta R \equiv \sqrt{(\Delta\eta)^{2} + (\Delta\phi)^{2}}\).}

The ATLAS detector~\cite{PERF-2007-01} at the LHC covers nearly the entire solid angle around the collision point.\footnote{\AtlasCoordFootnote}
It consists of an inner tracking detector surrounded by a thin superconducting solenoid, electromagnetic and hadronic calorimeters,
and a muon spectrometer incorporating three large superconducting air-core toroidal magnets.
 
The inner-detector system (ID) is immersed in a \qty{2}{\tesla} axial magnetic field
and provides charged-particle tracking in the range \(|\eta| < 2.5\).
The high-granularity silicon pixel detector covers the vertex region and typically provides four measurements per track,
the first hit generally being in the insertable B-layer (IBL) installed before Run~2~\cite{ATLAS-TDR-19,PIX-2018-001}.
It is followed by the SemiConductor Tracker (SCT), which usually provides eight measurements per track.
These silicon detectors are complemented by the transition radiation tracker (TRT),
which enables radially extended track reconstruction up to \(|\eta| = 2.0\).
The TRT also provides electron identification information
based on the fraction of hits (typically 30 in total) above a higher energy-deposit threshold corresponding to transition radiation.
 
The calorimeter system covers the pseudorapidity range \(|\eta| < 4.9\).
Within the region \(|\eta|< 3.2\), electromagnetic calorimetry is provided by barrel and
endcap high-granularity lead/liquid-argon (LAr) calorimeters,
with an additional thin LAr presampler covering \(|\eta| < 1.8\)
to correct for energy loss in material upstream of the calorimeters.
Hadronic calorimetry is provided by the steel/scintillator-tile calorimeter,
segmented into three barrel structures within \(|\eta| < 1.7\), and two copper/LAr hadronic endcap calorimeters.
The solid angle coverage is completed with forward copper/LAr and tungsten/LAr calorimeter modules
optimised for electromagnetic and hadronic energy measurements respectively.
 
The muon spectrometer (MS) comprises separate trigger and
high-precision tracking chambers measuring the deflection of muons in a magnetic field generated by the superconducting air-core toroidal magnets.
The field integral of the toroids ranges between \num{2.0} and \qty{6.0}{\tesla\metre}
across most of the detector.
Three layers of precision chambers, each consisting of layers of monitored drift tubes, cover the region \(|\eta| < 2.7\),
complemented by cathode-strip chambers in the forward region, where the background is highest.
The muon trigger system covers the range \(|\eta| < 2.4\) with resistive-plate chambers in the barrel, and thin-gap chambers in the endcap regions.
 
The luminosity is measured mainly by the LUCID--2~\cite{LUCID2} detector that records Cherenkov light produced in the quartz windows of photomultipliers located close to the beampipe.
 
Events are selected by the first-level trigger system implemented in custom hardware,
followed by selections made by algorithms implemented in software in the high-level trigger~\cite{TRIG-2016-01}.
The first-level trigger accepts events from the \qty{40}{\MHz} bunch crossings at a rate below \qty{100}{\kHz},
which the high-level trigger further reduces in order to record complete events to disk at about \qty{1}{\kHz}.
 
A software suite~\cite{SOFT-2022-02} is used in data simulation, in the reconstruction
and analysis of real and simulated data, in detector operations, and in the trigger and data acquisition
systems of the experiment.

\section{Data and simulated event samples }
\label{sec:datasetMC}

 
The results presented in this paper are based on data collected by the ATLAS experiment between 2015 and 2018 during Run 2 of the LHC.
The dataset corresponds to an integrated luminosity of 140 fb$^{-1}$~\cite{DAPR-2021-01,LUCID2} of \(pp\) collisions at a center-of-mass energy of 13~\TeV.
Events were selected at trigger level based on the presence of at least one electron~\cite{TRIG-2018-05} or muon~\cite{TRIG-2018-01}.
These single-lepton triggers were configured with \pT thresholds ranging from 20 to 26~\GeV, depending on the lepton flavor and data-taking period.
Every detector subsystem must have been in operation during data collection. Moreover, stringent data quality criteria were applied to ensure the reliability of the data~\cite{DAPR-2018-01}.
 
Simulated samples of SM processes are based on Monte Carlo (MC) techniques,
and the main contributing samples are described in the following.
All simulated samples are processed through the ATLAS simulation infrastructure~\cite{SOFT-2010-01}
including a detailed model of the ATLAS detector based on \GEANT~\cite{Agostinelli:2002hh}.
 
The effect of multiple interactions in the same and neighboring bunch crossings (\pileup) are modeled by overlaying the
simulated hard-scattering event with inelastic \(pp\) events generated with \PYTHIA~8.186~\cite{Sjostrand:2007gs} using the \NNPDF[2.3LO]~\cite{Ball:2012cx} set of parton distribution functions
(PDF) and the A3 set of tuned parameters~\cite{ATL-PHYS-PUB-2016-017}.
The MC events are weighted to reproduce the
distribution of the average number of interactions per bunch crossing
($\mu$) observed in the data, where the $\mu$
value in data is rescaled by a factor of \(1.03\pm 0.04\) to improve
agreement between data and simulation in the visible inelastic
proton--proton cross-section~\cite{STDM-2015-05}.
A further correction to the distribution is derived in the sample selected by this analysis such that the $\mu$ distribution in simulated events matches the one in data for the calibration measurement.

The analysis presented in this paper is designed to select \ttbar events where each of the top quarks decays leptonically, hence the modeling of this SM
process is particularly important. A suite of MC samples with different accuracy in the matrix element (ME) and different models for the parton shower (PS) are used for the \ttbar process.
The baseline \ttbar simulated sample was produced using the \powhegbox~\cite{Frixione:2007nw}~v2 generator with a ME at next-to-leading
order (NLO) in quantum chromodynamics (QCD) in the five flavor scheme (5FS), and interfaced to \PYTHIA~8.230~\cite{Sjostrand:2014zea} with the A14 set of tuned parameters~\cite{ATL-PHYS-PUB-2014-021} to model PS, hadronization, and the underlying event.
The PDF sets \NNPDF[3.0NNLO] and \NNPDF[2.3LO]~\cite{Ball:2014uwa,Ball:2012cx} were used in the ME-PS matching, respectively.
The \hdamp parameter, which controls the matching in \POWHEG and
effectively regulates the high-\pT radiation against which the
\ttbar system recoils, was set to 1.5\,$m_{t}$~\cite{ATL-PHYS-PUB-2016-020}.
The functional form of the renormalization and factorization scales was
set to the default scale ($\sqrt{m_t^{2} + p_{\textrm{T},t}^2}$).

Alternative \ttbar simulation samples were generated using \powhegbox~v2
interfaced to \herwigseven.04~\cite{Bellm:2015jjp} with the
H7-UE-MMHT set of tuned parameters.
Uncertainties related to initial- and final-state radiation (ISR, FSR) are estimated by reweighting the
baseline \ttbar events such that initial (final) parton shower radiation~\cite{ATL-PHYS-PUB-2016-020} is increased or reduced and by using an alternative \powhegbox~v2 + \PYTHIA~8.230 sample
with $h_{\textrm{damp}}$ set to $3m_{t}$ and
the parameter variation group \textsc{Var3} (described in Ref.~\cite{ATL-PHYS-PUB-2016-020}) increased, leading to increased ISR.

Samples with \ttbar + $b\bar{b}$ MEs were produced at NLO QCD accuracy with the \POWHEGBOXRES~\cite{Jezo:2018yaf}
generator and \OPENLOOPS~\cite{Buccioni:2019sur,Cascioli:2011va,Denner:2016kdg}, using a pre-release
of the implementation of this process in \POWHEGBOXRES provided by the authors~\cite{ttbbPowheg},
with the \NNPDF[3.0NLO]nf4~\cite{Ball:2014uwa} PDF set. It was interfaced with \PYTHIA[8.240]~\cite{Sjostrand:2014zea},
using the A14 set of tuned parameters~\cite{ATL-PHYS-PUB-2014-021} and the \NNPDF[2.3LO] PDF set.
The four-flavor scheme was used with the \(b\)-quark mass set to \qty{4.95}{\GeV}.
The factorization scale was set to $0.5\times\Sigma_{i=t,\bar{t},b,\bar{b},j}m_{\mathrm{T},i}$,
the renormalization scale was set to $\sqrt[4]{\vphantom{m_{b}^2} m_{\mathrm{T},t} \cdot m_{\mathrm{T},\bar{t}}\cdot m_{\mathrm{T},b} \cdot m_{\mathrm{T},\bar{b}}}$,
and the \hdamp parameter was set to \(0.5\times\Sigma_{i=t,\bar{t},b,\bar{b}}m_{\mathrm{T},i}\).
Events containing two $B$-hadrons produced in the decay of the two top quarks and either one or two additional $B$-hadrons with $\pt > 5\;\GeV$ and matched to a particle jet are vetoed in the nominal \ttbar\ sample and taken from the \ttbar + $b\bar{b}$ sample. Events with an additional $c$-flavored hadron with $\pt > 5\;\GeV$ and matched to a particle jet undergo the same procedure.

Additional processes entering the analysis selection arise from the production of a single top quark, either through the associated production with a $W$ boson ($tW$) or in $t$-channel and $s$-channel production. These contributions were modeled by the \powhegbox~v2~\cite{Frederix:2012dh,Alioli:2009je,Re:2010bp}
generator with ME at~NLO. For $t$-channel production, events were
generated in the four flavor scheme with the \NNPDF[3.0NLO]nf4
PDF set,
and the functional form of the renormalization and factorization scale
was set to $\sqrt{m_{b}^2 + p_{\textrm T,b}^2}$ following the
recommendation of Ref.~\cite{Frederix:2012dh}.
For $s$-channel and $tW$ production, events were generated in the five
flavor scheme with the \NNPDF[3.0NLO] PDF set, and the functional form of
the renormalization and factorization scale was set to the default
scale, which is equal to the top-quark mass.
For $tW$ production, the diagram removal scheme~\cite{Frixione:2008yi} was employed to treat the interference with \ttbar\ production~\cite{ATL-PHYS-PUB-2016-020}.
To evaluate the impact of the PS and hadronization model, the same
events produced for the nominal \powhegboxpythia\ samples were used,
but they were showered with \herwigseven.04.
To assess the uncertainty due to the choice of the matching scheme,
the nominal samples are compared with samples generated with the
\mgamc\ v2.6.2~\cite{Alwall:2014hca} generator at NLO in QCD, in the four flavor scheme for
$t$-channel production, and the five flavor scheme for $s$-channel and
$tW$ production. For $t$-channel production (respectively $s$-channel,
$tW$ production), the \NNPDF[3.0NLO] PDF set is used.
 
To evaluate the effect of employing the diagram subtraction scheme instead of the diagram removal scheme~\cite{Frixione:2008yi,Re:2010bp} for $tW$ events, alternative samples were produced,
where the diagram subtraction scheme was used.
All single-top-quark samples were showered with \PYTHIA~8.230, unless otherwise stated for studies assessing systematic uncertainties.

Additional small contributions are expected from \Zjets processes, which were modeled with the \SHERPA[2.2.1]~\cite{Bothmann:2019yzt} generator
providing NLO MEs for up to two partons in addition to the $Z$ boson and leading-order (LO) MEs for up to four partons calculated with the Comix~\cite{Gleisberg:2008fv} and \OPENLOOPS libraries. These samples were matched with the \SHERPA parton shower~\cite{Schumann:2007mg} using the \MEPSatNLO prescription~\cite{Hoeche:2011fd,Hoeche:2012yf,Catani:2001cc,Hoeche:2009rj}. The \NNPDF[3.0NNLO] set of PDFs~\cite{Ball:2014uwa} was used.
 
Several rare processes such as $t\bar{t}V$ (where $V$ is either a $W$ or $Z$ boson), $t\bar{t}H$, four-top-quark production and $VV$ are considered as well, however, they have a negligible contribution in the phase space considered.
 
The \EVTGEN[1.6.0] program ~\cite{Lange:2001uf} was employed to simulate the
decay of $b$- and $c$-flavored hadrons in all samples except those
generated using \SHERPA, for which the default configuration recommended
by the \SHERPA authors was used.


\section{Object reconstruction}
\label{sec:objects}

 
Tracks are reconstructed from hits inside the inner tracking detector~\cite{ATL-PHYS-PUB-2015-051,PERF-2015-08}. In addition to acceptance criteria of $|\eta| < 2.5$ and $\pt > 500\ \MeV$, the tracks must satisfy a set of quality criteria defined by the \textit{Loose} working point in Ref~\cite{ATL-PHYS-PUB-2015-051}. Tracks should have at least seven hits in the silicon detectors with no more than one detector module shared with other tracks, no more than one missing hit in the pixel detector, and no more than two missing hits in the SCT detector.
Events are required to contain at least one vertex with two or more associated tracks, and the vertex with the highest $p_{\text{T}}^2$ sum of the associated tracks is taken as the primary vertex (PV)~\cite{ATL-PHYS-PUB-2015-026}.
Transverse and longitudinal impact parameters,  $d_0$ and $z_0$, are used to identify tracks that originate at a certain displacement from the PV (e.g. $B$-hadron  decay products). The transverse impact parameter, $d_0$, is the point of closest approach to the measured beam line position; and $z_0$, required to satisfy $|z_0 \sin \theta | < 3$ mm, is the difference between the longitudinal position of the track along the beam line at the point where $d_0$ is measured and the longitudinal position of the primary vertex~\cite{Billoir:1992yq}. The latter requirement reduces the contamination from \pileup, secondary and fake tracks.
The collection of tracks satisfying the criteria stated above is used as input to the TC-LVT tagger for the reconstruction of soft secondary vertex (SSV) candidates.
The complete set of requirements for tracks used for standard high-$\pt$ $b$-tagging algorithms are described in Ref.~\cite{FTAG-2019-07}.
 
Electron candidates are reconstructed from energy deposits (clusters) in the electromagnetic calorimeter associated with reconstructed tracks in the inner detector~\cite{EGAM-2018-01}.
Candidates that satisfy $\pt>10~\GeV$ and $|\eta|<2.47$ are selected, excluding the calorimeter transition region $1.37<|\eta|<1.52$.  Electrons must satisfy the \textit{Tight} likelihood-based identification criterion and are required to satisfy the criteria of the \textit{PLVLoose}
working point of the `Prompt Lepton Tagger'~\cite{HIGG-2017-02}, a multivariate isolation discriminant used to reject non-prompt leptons from heavy-flavor decays.
They are further required to have $|z_0\sin{\theta}| < 0.5$~mm and $|d_0/\sigma(d_0)|<5$, where the longitudinal and transverse impact parameters $d_0$ and $z_0$ are computed relative to the primary vertex and $\sigma(d_0)$ is the resolution of the track transverse impact parameter $d_0$. It is computed using an iterative Gaussian fit performed on the core of the impact parameter distribution~\cite{Magliocca:2791559}.
 
Muon candidates are reconstructed from track segments in the various layers of the muon spectrometer, and matched with tracks from the inner detector~\cite{MUON-2018-03}.
The final muon candidates are refitted using the complete track information from both detector systems, and required to satisfy $\pt>10~\GeV$ and $|\eta|<2.5$. Muons are required to satisfy the \textit{Medium} quality requirements and to satisfy the isolation criteria of the \textit{PLVLoose} working point of the Prompt Lepton Tagger. The absolute value of the muon $d_0$ significance must be less than 3, and the value of $|z_0\sin{\theta}|$ must be less than 0.5~mm.
 
Jets are reconstructed by clustering particle-flow objects~\cite{PERF-2015-09} with the anti-\kt algorithm~\cite{Cacciari:2008gp,Fastjet} with a radius parameter of $R=0.4$ and a four-momentum recombination scheme.
The jet energy is corrected to the particle level by the application
of a jet energy scale calibration derived from \pp collision data
and simulation at $\sqrt{s}=$~13~\TeV~\cite{JETM-2018-05}.
Baseline jets are required to have \pt$>20~\GeV$ and $|\eta| < 2.5$.
They are also required to satisfy a \textit{Tight} \pileup rejection criterion based on the jet vertex tagger (JVT) score~\cite{PERF-2014-03}: for jet \pt $\in [20,60]~\GeV$ and $|\eta|<2.4$, the JVT score must be higher than 0.5.
 
The identification of \bjets with $\pt > 20\; \GeV$ is done using the DL1r algorithm~\cite{FTAG-2019-07}, a high-level tagger based on a deep neural network (NN). The NN is trained on simulated \ttbar{} and $Z'$ events and uses the results of specialized low-level taggers, based on track impact parameters, secondary vertices and decay topologies, to obtain the probability for each jet to be a $b$-, $c$- or light-jet.
In this measurement the 85\% working point is used, corresponding to an 85\% efficiency to correctly tag a $b$-jet in simulated \ttbar\ events.

To avoid double counting of the objects used in the analysis, special care is taken to remove any potential overlap. This procedure is based on two object distances defined by $\Delta R_{y}=\sqrt{(\Delta y)^2+(\Delta \phi)^2}$, where $y$ is the rapidity and $\phi$ the azimuthal angle of each object.
The removal is carried out systematically
starting with electrons, where the closest jet within  $\Delta R_{y}=0.2$ of a selected electron is removed.
If the nearest jet surviving that selection is within $\Delta R_{y} = 0.4$ of the electron, the electron is discarded.
Muons are removed if they are separated from the nearest jet by $\Delta R_{y} < 0.4$, which reduces the background from heavy-flavor decays inside jets.
However, the muon is kept and the jet removed instead, if fewer than three tracks are associated with the jet. This criterion avoids inefficiency for the selection of high-energy muons undergoing significant energy loss in the calorimeter.
 
The missing transverse momentum, with magnitude \met, is defined as the negative vector sum of the transverse momenta of all selected and calibrated electrons, muons and jets in the event, with an extra term added to
account for energy in the event that is not associated with any of these objects~\cite{JETM-2020-03}.
 
\section{Soft secondary vertex algorithm definition and performance}
\label{sec:tclvtdefperf}
\subsection{TC-LVT algorithm description}
 
The TC-LVT algorithm is based on vertexing techniques first developed in the context of standard $b$-tagging, which are retuned and applied outside hadronic jets.
It aims to identify $B$-hadrons outside hadronic jets by reconstructing displaced secondary vertices.
The key ingredients of the algorithm are  seed tracks, cluster tracks (which are inputs to track clusters in the following) and reconstructed vertices as defined by the selection criteria summarized in Table~\ref{tab:tc-lvt WP}.
These criteria define the \textit{Loose} working point and were optimized to maximize the efficiency of the algorithm.
 
As a first step a set of \textit{seed tracks} is extracted. The seed tracks are characterized by a high transverse momentum of $\pT > 1.5\,$~\GeV and $|d_0 / \sigma(d_0)| > 0.5$.
Tracks that are also matched to a track-jet~\cite{ATL-PHYS-PUB-2014-013} with $\pT > 20$~\GeV are not considered. The list of seed tracks is ordered in decreasing \pt.
Clusters of tracks are built around these seed tracks by adding additional high-displacement tracks ($|d_0 / \sigma(d_0)|$  $>$ 1.5) within an angular distance to the seed track of $\Delta R^{\mathrm{track}}_{\mathrm{seed}} < 0.75$ and within a track-to-track distance of closest approach $d^{\mathrm{track}}_{\mathrm{seed}} < 0.25\,$mm.
Once tracks are associated with a cluster, they are not considered further for other clusters.
 
The next step of the algorithm is the vertexing process, which is accomplished using the Single Secondary Vertex Finding (\verb!SSVF!) algorithm~\cite{ATL-PHYS-PUB-2017-011}. For clarity, in the following the main features of this algorithm are described and the reader is refered to Ref.~\cite{ATL-PHYS-PUB-2017-011} for details. For each identified cluster, the \verb!SSVF! algorithm is executed on all tracks within  $\Delta R = 0.4$ from the vector sum of the momenta of all tracks associated with the cluster.
First, all the possible pairs of tracks originating from nearby points are combined into two-track vertices. Further requirements are applied to the two-track vertices to reduce the number of fake vertices and material interactions. For example, the tracks associated with the vertex must not have any hits in detector layers at a radius smaller than that of the vertex itself. Two-track vertices located in regions of the detector with large amounts of material are eliminated according to the procedure described in Ref.~\cite{ATL-PHYS-PUB-2017-011}. Two-track vertices with invariant masses consistent with $K^0_S$ and $\Lambda^0$ decays are identified, and the corresponding tracks are removed if the impact parameter of the reconstructed pair is consistent with the primary vertex. Similarly, two-track vertices with invariant masses below 40~\MeV under the $e^+ e^-$ hypothesis are excluded to reject photon conversions.
Finally, all tracks contributing to the cleaned set of two-track vertices are  combined into one list of selected tracks. This track list is supplied to the vertex fitter.
 
The vertex fitter runs iteratively on all the tracks in the list, trying to fit one secondary vertex from all these tracks. In each iteration the track with the largest $\chi^2$(track, vertex) of the track-vertex association is removed and the vertex fit is repeated until an acceptable vertex $\chi^2$(vertex) is found. The mass \mvtx of the vertex is  defined as the invariant mass of the four-vector sum of the constituent tracks and required to be above 600~\MeV and below 6~\GeV. The higher threshold is chosen to reject random track crossings while maximizing efficiency for true $B$-hadron vertices. The lower threshold is optimized to reject the most common light-flavored long-lived hadrons.
For similar reasons, the \ptvtx of the reconstructed vertex is defined as the \pT  of the four-vector sum of the constituent tracks, and required to be above 3~\GeV.
 
\begin{table}[h]
\centering
\renewcommand{\arraystretch}{1.25}
\caption{Selection criteria applied to the objects in the vertex reconstruction of the TC-LVT algorithm.}
\begin{tabular}{|l|l|l|}
\toprule
Seed track  ~~~~~~~~~~~~~~~~~~~~&  Cluster tracks  ~~~~~~~~~~~~~~~~~&  Vertex ~~~~~~~~~~~~~ \\ \midrule
$|d_0 / \sigma(d_0)|$  $>$ 0.5  &  $|d_0 / \sigma(d_0)|$  $>$ 1.5  &  600$\,$\MeV  $<$  \mvtx  $<$  6$\,$\GeV  \\
\pT  $>$ 1.5$\,$ \GeV  &  $\Delta R^{\mathrm{track}}_{\mathrm{seed}}$  $<$ 0.75  &  \ptvtx  $>$ 3$\,$ \GeV \\
 
&  $d^{\mathrm{track}}_{\mathrm{seed}}$  $<$ 0.25$\,$mm & \\
\bottomrule
\end{tabular}
 
\label{tab:tc-lvt WP}
\end{table}
 
As a last step, if two vertices are found they are merged if their angular distance $\Delta R < 0.3$.  This occurs rarely but helps the treatment of secondary or tertiary decays in the same hadron decay chain. The four-vectors of the merged vertices are added and the resulting vertex retains the parameter values, e.g. \Lxy (the displacement in the transverse plane of the secondary vertex relative to the primary vertex) or $L_{3d}$ (the significance of the 3D distance relative to the primary vertex), of the merged vertex with smallest \Lxy.

Objects identified with the TC-LVT algorithm, henceforth referred to as soft secondary vertices or SSVs, are removed if they are within $\Delta R <$ 0.2 of a reconstructed lepton or within $\Delta R <$ 0.6 of a reconstructed jet.
 
\subsection{TC-LVT algorithm performance}
 
The performance of the TC-LVT algorithm is investigated using the full set of samples described in Section~\ref{sec:datasetMC} and applying the event selection described in Section~\ref{sec:measurement}. The parameters used in this study are:
 
\begin{itemize}
\item $\varepsilon_{\text{ssv}}$: probability that a true \ssv\ is identified by the \tclvt\ tagger, and
\item $n_{f}$: expected average number of fake \ssv s per event.
\end{itemize}
The \textit{Loose} working point of the TC-LVT algorithm (see Table~\ref{tab:tc-lvt WP}) is used, which is designed to maximize the efficiency of the algorithm while achieving the lowest possible fake rate.
The classification of true and fake \ssv s is done based on a cone-matching requirement between the reconstructed \ssv\ and a simulated $B$-hadron in acceptance. A $B$-hadron is considered to be in acceptance if it meets the criteria of $\pt > 2~\GeV$, $|\eta| < 2.8$, and $\Delta R_{ \mathrm{min}}(B, \mathrm{jet})>$ 0.3, where the latter ensures that the $B$-hadron is located outside of any reconstructed jet.
These acceptance requirements are driven by the ATLAS tracking detector geometry and the typical jet distance parameters used in the jet reconstruction.
SSVs fulfilling $\Delta R(B, \ssv) < 0.3$ are considered true SSVs, and their event multiplicity is indicated by $\mathcal{N}_{\text{ssv}}$. Unmatched SSVs are considered fake SSVs.
\ssv s matched to a $c$-flavored hadron in acceptance are also categorized as true \ssv s, as long as the $c$-flavored hadron is separated from any $B$-hadron in acceptance by $\Delta R > 0.3$. Events of this kind represent a minor contribution in the total calibration sample.
 
\begin{figure}[tb]
\centering
\begin{subfigure}{.36\textwidth}
\includegraphics[width=0.95\textwidth]{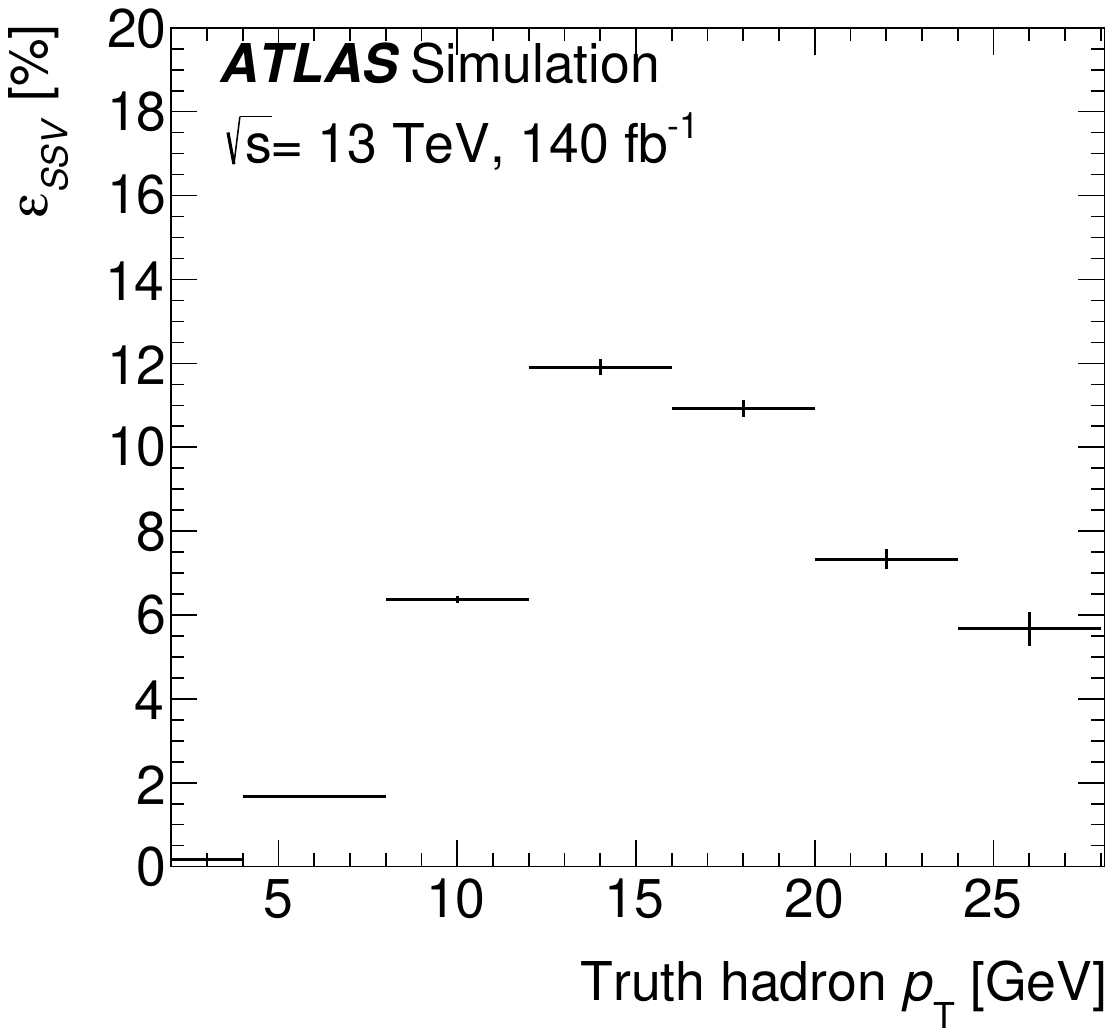}
\caption{\label{fig:efficiency_pT}}
\end{subfigure}\\
\begin{subfigure}{.36\textwidth}
\includegraphics[width=0.95\textwidth]{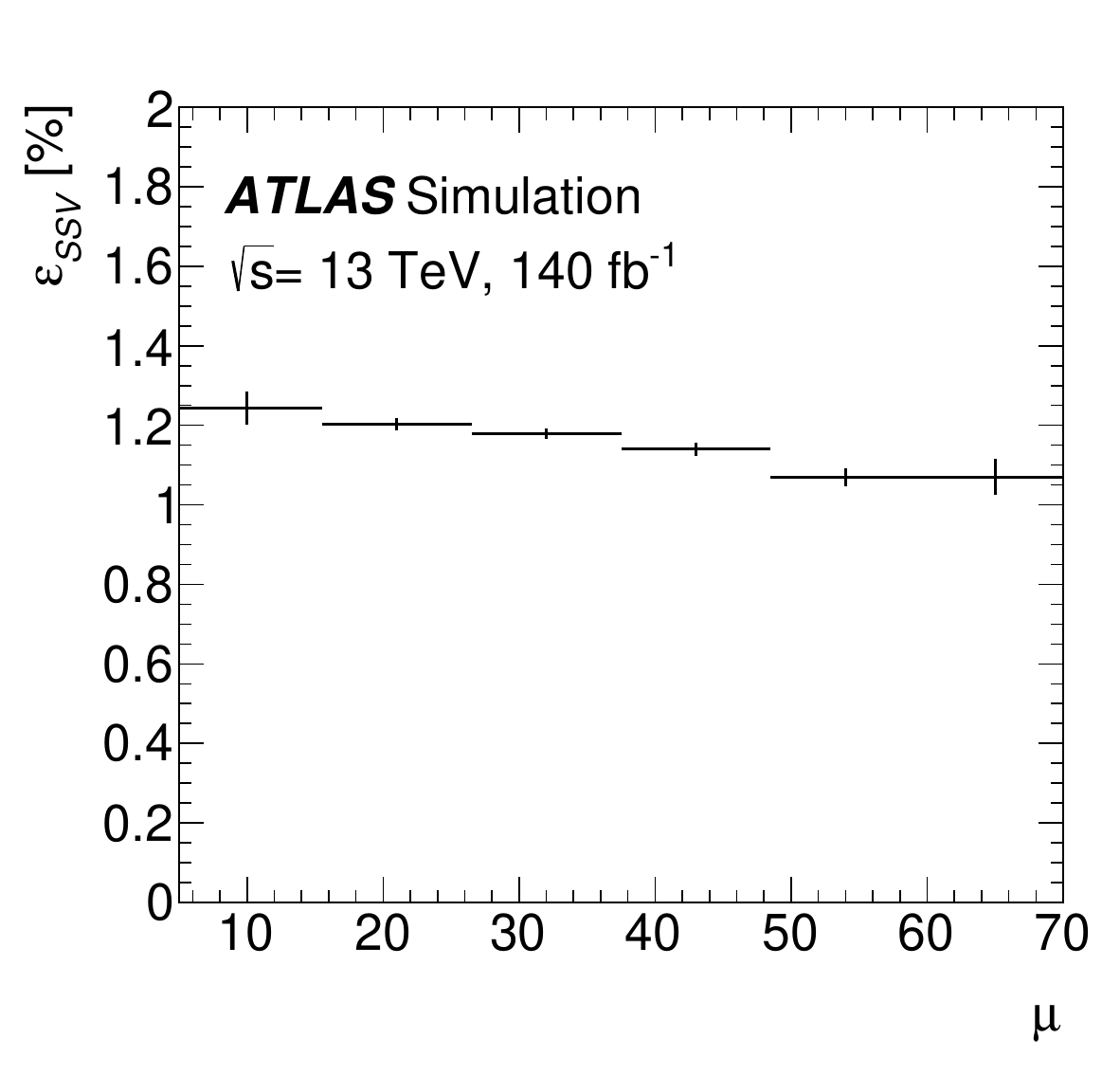}
\caption{\label{fig:efficiency_mu}}
\end{subfigure}
\begin{subfigure}{.36\textwidth}
\centering
\includegraphics[width=0.95\textwidth]{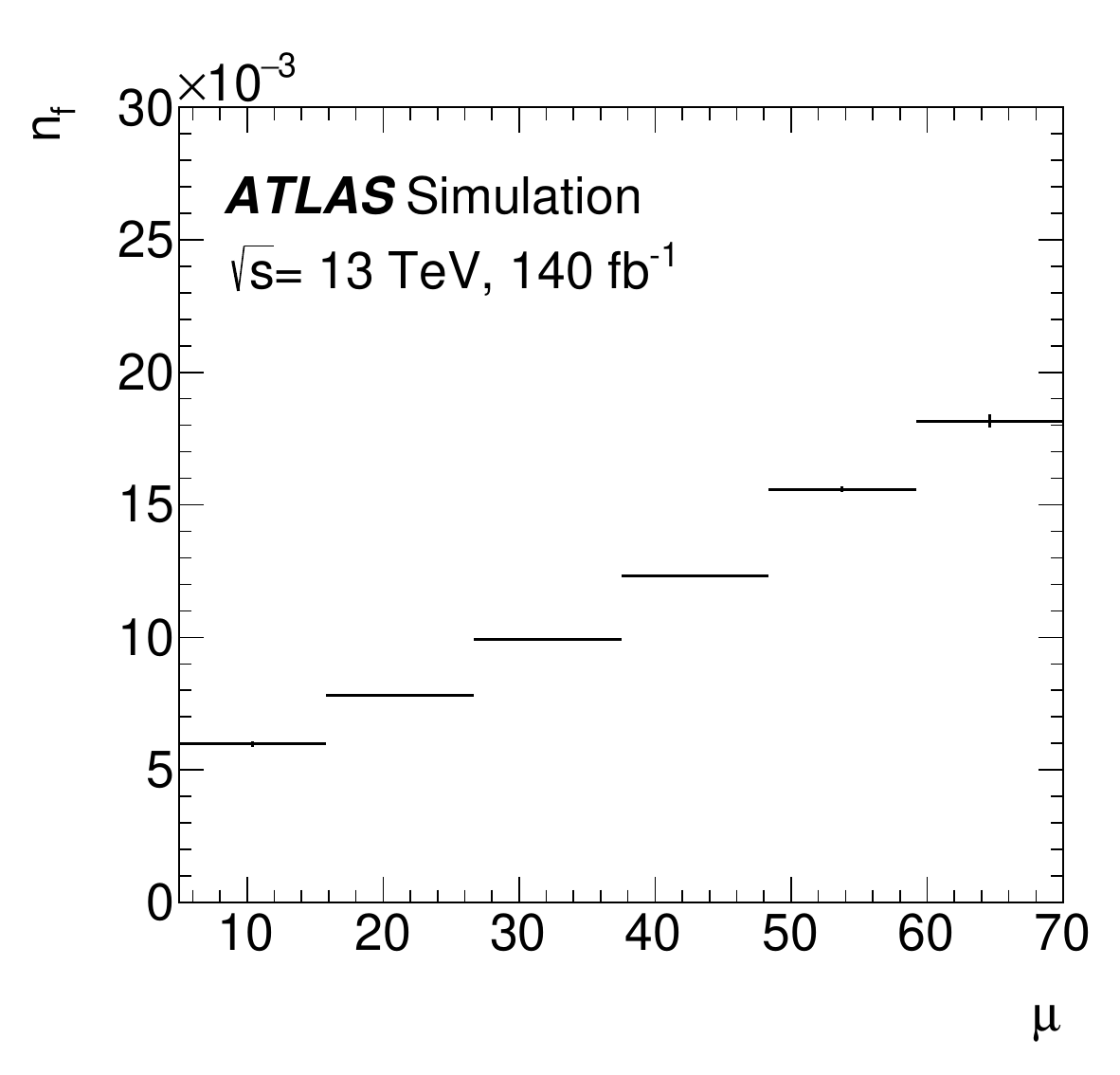}
\caption{\label{fig:fakerate_mu}}
\end{subfigure}
\caption{TC-LVT efficiency to reconstruct true \ssv s as a function of (a) truth hadron \pT  and (b) average number of interactions per bunch crossing, $\mu$. (c) Average number of fake \ssv s per event, $n_{f}$, as a function of $\mu$.
A $B$-hadron is considered to be in acceptance if it meets the criteria of $\pt > 2~\GeV$, $|\eta| < 2.8$, and $\Delta R_{ \mathrm{min}}(B, \mathrm{jet})>$ 0.3, where the latter ensures that the $B$-hadron is located outside of any reconstructed jet. \ssv s matched to a $c$-flavored hadron in acceptance are also categorized as true \ssv s, as long as the $c$-flavored hadrons is separated from any $B$-hadron in acceptance by $\Delta R > 0.3$.
The performance is evaluated on all SM processes. Events are selected in a phase space enriched with di-leptonic \ttbar\ events (see Section~\ref{sec:measurement} for details). }
\end{figure}

Figure~\ref{fig:efficiency_pT} shows the TC-LVT efficiency to reconstruct true SSVs as a function of the generator-level (`truth’) hadron \pT. The efficiency first increases as a function of $B$-hadron \pt as the displacement of secondary vertices also increases. The turnover at about $15\;\GeV$ is due to the fact that a $B$-hadron of such \pt is more likely to be reconstructed as a jet instead.
Figure~\ref{fig:efficiency_mu} shows the TC-LVT efficiency to reconstruct a true SSV as a function of the average number of interactions per bunch crossing ($\mu$). While the TC-LVT efficiency is rather independent of $\mu$, Figure~\ref{fig:fakerate_mu} shows that the average number of fake \ssv s per event ($n_{f}$)  is increasing with $\mu$.
This trend is expected from the fact that most of the fake \ssv s originate from random track crossings. The fake rate calibration is therefore carried out in two distinct $\mu$ regions.
 
\FloatBarrier
 
\section{Calibration of the soft secondary vertex tagger}
\label{sec:analysis}
\subsection{Measurement strategy}
\label{sec:measurement}

To account for the difference of the \tclvt\  performance in data with respect to the MC simulation, correction scale factors (SFs) are estimated by performing the calibration procedure following the strategy described in this section.
The calibration is carried out in a sample enriched in \ttbar\ events, which overall is characterized by low to medium \MET.
Three SFs are
extracted, as parameters of interest (POIs), with a simultaneous likelihood fit in multiple event categories (fit regions) enriched in true and fake \ssv s. These are the efficiency scale factor (\SFeff) \footnote{
It was verified that fitting two separate efficiency scale factors from the low and high \pileup\ data would yield compatible results.
} and two fake-rate scale factors (\SFfakelow, \SFfakehigh).  To account for track density effects on the fake rate (Figure~\ref{fig:fakerate_mu}), two separate \SFfake parameters are determined from low and high \pileup\ data ($\mu_\text{L}$, $\mu_\text{H}$), respectively.
The applicability of these correction factors to simulated events with high  \MET, where many SUSY searches are carried out, is validated in dedicated regions.

Events are selected based on the expected topology of the \ttbar process, where both top quarks decay leptonically. The events are required to contain exactly two leptons ($\ell$, either electrons or muons) of different flavor and opposite charge, where the leading \pT lepton is required to have
$\pT > 27$~\GeV and the sub-leading lepton $\pT > 10$~\GeV. The invariant mass of the two leptons $m_{\ell \ell}$ must be $>50$~\GeV.
Additionally, events are required to have at least one \btagged jet with a  $\pT >  20$~\GeV.

Events are further categorized based on the number of reconstructed $b$-jets, $\mathcal{N}_{b}$.
Di-leptonic \ttbar events typically contain two true $B$-hadrons in the event, given that the branching ratio of top quarks decaying to a $W$ boson and a $b$-quark is nearly 100\%.
Separating \ttbar events with either exactly one or two reconstructed $b$-jets, for brevity referred to as $1b$ and $2b$, provide regions that are enhanced in true and fake SSVs, respectively. While \ttbar events dominate the $2b$ region with about 96\% purity, the $1b$ region has a contribution from single-top-quark processes of about 10\%.
However, the influence of non-\ttbar processes in the calibration sample is negligible.
 
In the $1b$ region, one $b$-quark gives rise to a \bjet that is identified as such by the DL1r algorithm, while the second $b$-quark, unless missed by the \btagging\ algorithm, does not give rise to a \bjet. Instead, it leads to a $B$-hadron that is reconstructed as a SSV. In the $2b$ region, the two $b$-quarks originating from the top-quark decay are already reconstructed as \bjets. The region is therefore depleted in non-reconstructed  $B$-hadrons and most reconstructed SSVs in that region are fakes (except for events arising from \ttbar + heavy flavor (HF)).

Events are further categorized by reconstructed \ssv\ multiplicity, as well as into low \pileup events by requiring $\mu_{\text{L}} \in [0, 30)$ and high \pileup events by requiring $\mu_{\text{H}} \in [30, 80]$, such that each region has a similar number of events. The average $\mu$ value of the events selected in the low \pileup region is $22$, while the average $\mu$ value in the high \pileup region is $42$.
Events with $\mu > 80$ represent a negligible fraction of the total number of events and are excluded from the fit regions.
The region requiring two \bjets and two \ssv s exhibits a yield that is too low, and is therefore excluded from the fit. Only events with $\met < 150\;\GeV$ are retained for the calibration measurement regions.
 
For each calibration region with zero or one \ssv, a validation region is defined satisfying the same event selection criteria except for the \met requirement, which is reversed: $\met \geq 150\;\GeV$. A validation region with two \ssv s is not defined since the event yield is too low.
These regions validate the scale factors in an independent sample and also aim to show that the calibration results can be used in the more extreme phase space typical of many SUSY searches as well.
Events with zero \ssv s are used to constrain the overall \ttbar\ normalization and provide a better control over systematic uncertainties associated with the modeling of the process.
Among these, the $2b$ regions show the strongest constraining power due to their higher purity in \ttbar\ events.
Two independent normalization factors ($\mu_{\ttbar}$ and $\mu_{\ttbar}^{\text{high-\met}}$) are employed for the calibration regions and the  $\text{high}$-$\met$ validation regions to account for an imperfect modeling of the \ttbar process.
The `$0\text{SSV}\;2b,\; \text{high}$-$\met$' region  is the only high-\met region used in the fit. It is inclusive in \pileup and it constrains only the $\mu_{\ttbar}^{\text{high-\met}}$ parameter. Hence it provides no constraints on the calibration POIs.
A summary of all the regions used in the calibration measurement is given in Table \ref{tab:tc-lvt fit regions}.
 
\begin{table}
\centering
\caption{Summary of all the selections used for the calibration measurement and its validation. The requirement on \ssv\ multiplicity ($\mathcal{N}_{\ssv}$), \bjet multiplicity ($\mathcal{N}_b$), average number of interactions per bunch crossing ($\mu$) and \met are indicated in the table. The last column indicates which parameters of the fit are constrained or validated by each selection. }
\label{tab:tc-lvt fit regions}
\begin{tabular}{|l|c|c|c|c|m{35mm}|}
\toprule
Region label & $\mathcal{N}_{\textrm{\ssv}}$ & $\mathcal{N}_b\;$ & $\mu\;$  & \met & Purpose\\\midrule
0SSV 2b,  $\muactual_{\text{L}}$          & 0 & 2 & $[0, 30)$ & $< 150$    & constrain $\mu_{\ttbar}$ \\[1ex]
0SSV 2b,  $\muactual_{\text{H}}$          & 0 & 2 & $[30, 80]$ & $< 150$    & constrain $\mu_{\ttbar}$ \\[1ex]
0SSV 2b, high-\scriptsize{\met} & 0 & 2 & - & $\geq 150$ & constrain $\mu_{\ttbar}^{\text{high}-\met}$ \\
\midrule
1SSV 1b, $\muactual_{\text{L}}$ & 1 & 1 & $[0, 30)$ & $< 150$    & constrain \SFeff, \SFfakelow \\[1ex]
1SSV 1b, $\muactual_{\text{H}}$ & 1 & 1 & $[30, 80]$ & $< 150$    & constrain \SFeff, \SFfakehigh \\[1ex]
1SSV 1b, $\muactual_{\text{L}}$, high-\scriptsize{\met} & 1 & 1 & $[0, 30)$ & $\geq 150$    & validate \SFeff, \SFfakelow \\[1ex]
1SSV 1b, $\muactual_{\text{H}}$, high-\scriptsize{\met} & 1 & 1 & $[30, 80]$ & $\geq 150$    & validate \SFeff, \SFfakehigh \\
\midrule
1SSV 2b, $\muactual_{\text{L}}$ & 1 & 2 & $[0, 30)$ & $< 150$    & constrain \SFeff, \SFfakelow \\[1ex]
1SSV 2b, $\muactual_{\text{H}}$ & 1 & 2 & $[30, 80]$ & $< 150$    & constrain \SFeff, \SFfakehigh \\[1ex]
1SSV 2b, $\muactual_{\text{L}}$, high-\scriptsize{\met} & 1 & 2 & $[0, 30)$ & $\geq 150$    & validate \SFeff, \SFfakelow \\[1ex]
1SSV 2b, $\muactual_{\text{H}}$, high-\scriptsize{\met} & 1 & 2 & $[30, 80]$ & $\geq 150$    & validate \SFeff, \SFfakehigh \\
\midrule
2SSV 1b, $\muactual_{\text{L}}$ & 2 & 1 & $[0, 30)$ & $< 150$    & constrain \SFeff, \SFfakelow \\[1ex]
2SSV 1b, $\muactual_{\text{H}}$ & 2 & 1 & $[30, 80]$ & $< 150$    & constrain \SFeff, \SFfakehigh \\[1ex]
\bottomrule
\end{tabular}
 
\end{table}

In each fit region, simulated events are split according to the number of true \ssv s and the number of non-reconstructed $B$-hadrons in acceptance,
i.e. $B$-hadrons not matched to a SSV reconstructed by the TC-LVT and therefore represent \textit{missed} \ssv s.
Each simulated event acquires a weight composed of a tagging efficiency correction ($\mathcal{P}^\text{eff}$), a tagging inefficiency correction ($\mathcal{P}^\text{ineff}$) and a fake global event correction ($\mathcal{P}^\text{fake}$). These corrections are a function of the number of true and missed \ssv s, $\mathcal{N_{\text{true}}}$ and $\mathcal{N_{\text{miss}}}$ respectively, the efficiency in simulation, $\varepsilon_{\text{\ssv}}$, the expected number of fake SSVs per event from simulation $n_{f}$ and the SF POIs as stated in Eq.~\eqref{fitweight}:
 
\begin{align}
&w(\mathcal{N_{\text{true}}}, \mathcal{N_{\text{miss}}}, \mathcal{N_{\text{\ssv}}}) = \mathcal{P}^\text{eff} \ \times \ \mathcal{P}^\text{ineff} \ \times \ \mathcal{P}^\text{fake}  \notag \\
&= \prod_{1}^\mathcal{N_{\text{true}}}  \SFeff \ \times \ \prod_{1}^\mathcal{N_{\text{miss}}} \frac{1 \ - \SFeff \cdot \varepsilon_{\text{ssv}}(\mathcal{N}_{b})}{1 \ - \ \varepsilon_{\text{ssv}}(\mathcal{N}_{b})}
\times \prod_{\mu = \mulow, \muhigh} \frac{\textrm{Poiss}\left( \mathcal{N}_{\text{ssv}} -  \mathcal{N_{\text{true}}} | \ \SFfake(\mu) \cdot n_{f}(\muactual, \mathcal{N}_{b}) \right) }{\textrm{Poiss}\left( \mathcal{N}_{\text{ssv}} -  \mathcal{N_{\text{true}}} \ | \ n_{f}(\mu, \mathcal{N}_{b})\right)}. \label{fitweight}
\end{align}

The SF POIs are common for the $1b$ and the $2b$ selections and inclusive in $B$-hadron \pt.
The last term of the equation describes the SF correction to the fake rate as a ratio of conditional Poisson probabilities. This is calculated, globally per event, as the conditional Poisson probability of observing the number of fake reconstructed \ssv s in the event ($\mathcal{N}_{\text{ssv}} -  \mathcal{N_{\text{true}}}$) given the average number of fakes SSVs predicted by the simulation ($n_{f}(\muactual, \mathcal{N}_{b})$), where $ \mathcal{N}_{b}$ is the number of reconstructed $b$-jets. In the numerator of this term, $n_{f}$ is corrected by \SFfake, which is one of the free parameters of interest in the fit. Products over $\mathcal{N_{\text{true}}}$ and $\mathcal{N_{\text{miss}}}$ are truncated at four, i.e. consider only events with up to four $B$-hadrons in acceptance.
 
\begin{table}[h]
\centering
\caption{Input values to weight calculation of Eq.~\eqref{fitweight} obtained from simulated events. Relative statistical uncertainties amount to less than 1\% in all cases.  }
\label{montecarloweightinputs}
\begin{tabular}{|l|l|l|}
\toprule
Parameter & $\mathcal{N}_{b}$ = 1 & $\mathcal{N}_{b}$ = 2 \\ \midrule
$\varepsilon_{\text{ssv}}$ & 0.0173 & 0.0116 \\
$n_{f}(\mulow)$ & 0.0105 & 0.0078 \\
$n_{f}(\muhigh)$ & 0.0149 & 0.0125 \\
\bottomrule
\end{tabular}
 
\end{table}
 
Events with higher $B$-hadron multiplicities are negligible. The parameters $\varepsilon_{\text{\ssv}}$ and $n_{f}$ are estimated in simulated $\ttbar$ events as a function of $\mathcal{N}_{b}$ (and in two bins of $\mu$ for $n_{f}$) and provided as an input to the weight calculation. The values used are shown in Table~\ref{montecarloweightinputs}.   The performance difference between the $1b$ and the $2b$ selections is due to differences between the SSV momentum distribution in the two regions. These parameters were also estimated by using \Herwig as alternative parton shower modeling generator. It was found that $n_f$ does not depend on the shower generator choice. The efficiency, $\varepsilon_{\text{\ssv}}$, is also very similar between the two parton shower generators for $B$-hadron \pt smaller than $10\;\GeV$, while it is about 15\% smaller for the alternative shower simulation for larger $B$-hadron \pt. This difference is also considered in the systematic uncertainties of the measurement (Section~\ref{sec:systematics}).


\subsection{Systematic uncertainties}
\label{sec:systematics}


Various sources of uncertainties related to the underlying theoretical models for simulated processes as well as detector effects are considered in the measurement of the scale factors and summarized in this section. The systematic uncertainties are introduced to the likelihood function in the form of Gaussian-constrained Nuisance Parameters (NPs) $\theta_i$ and fitted simultaneously with the POIs, as introduced in the previous section.
 
Systematic uncertainties related to the track selection efficiency are determined through dedicated simulated samples where the amount of tracker material and the physical models were altered in the \GEANT implementation of the detector. Residual alignment uncertainties between data and simulation on the track parameters, including the transverse ($d_0$) and longitudinal ($z_0$) impact parameters and the track sagitta, as well as uncertainties on the impact parameter resolution are considered.
These effects, along with systematic variations of the number of fake tracks, are applied to the tracks used in the TC-LVT reconstruction algorithm, which is re-executed for each source of uncertainty
such that the number of true and fake \ssv\ in each event is re-evaluated.

A variation in the \pileup reweighting of simulated events is included to cover the uncertainty on the ratio of the predicted
inelastic cross-section in simulations compared to the one measured in data.
 
Uncertainties related to the energy scale (JES) and resolution~\cite{JETM-2018-05} of hadronic jets are derived by combining information from test-beam data, LHC collision data and simulation. Uncertainties related to the jet energy resolution have negligible impact on the measurement. Among the sources of uncertainty contributing to the JES, the most important uncertainties for this measurements are those related to the modeling of the \pileup, flavor composition of the jet calibration sample, and the different response for each jet flavor (see Ref.~\cite{JETM-2018-05} for details).
In addition, uncertainties related to the differences between data and simulation of the performance of the JVT algorithm~\cite{PERF-2014-03} are accounted for.

Dedicated measurements are carried out to assess the performance of the DL1r $b$-tagging algorithm on $b$-jets~\cite{FTAG-2018-01} and the mistagging performance for $c$-jets~\cite{ATLAS:2021cxe} and light-jets~\cite{ATLAS:2023lwk}.
These measurements derive corrections from data control samples, which are applied to account for
differences between data and simulation in the efficiency and mistag rate of the \btagging\ algorithm. The corrections used in this analysis are derived in pseudo-continuous bins of the tagger discriminant distribution~\cite{FTAG-2018-01}.
A large number of uncertainties are considered. A principal component analysis is performed on these uncertainties, and the resulting uncertainties are taken as uncorrelated components for the measurement described here.
An uncertainty in the relative fraction of quark versus gluon jets of 50\% is also applied to all simulated samples.

The combined  2015--2018 integrated luminosity, which is obtained using the LUCID-2 detector for the primary  measurement and  complemented by measurements using the inner detector and calorimeters, has an uncertainty of 0.83\% \cite{DAPR-2021-01}.
 
The reconstruction, identification, and isolation efficiencies of electrons and muons, as well as the efficiency of the trigger used to record the events, differ slightly between data and simulation. These differences are compensated for by dedicated SFs, which are measured using tag-and-probe techniques on $Z\to \ell^+\ell^-$ experimental and simulated samples~\cite{EGAM-2018-01,MUON-2018-03}. The SFs, with uncertainties, are applied as corrections to the simulated event weights. Additional sources of uncertainty arise from the corrections used to adjust the lepton momentum scale and resolution, which are also propagated to the simulated samples. All lepton-related uncertainties have an impact on the measurement smaller than 0.5\%.
 
Uncertainties related to the modeling of the \ttbar process include effects associated with missing higher order corrections in the perturbative expansion of the partonic cross-section and the PDF uncertainties.
The first point is addressed by varying the renormalization ($\mu_\textrm{r}$) and factorization ($\mu_\textrm{f}$) scales, while the latter is evaluated following the PDF4LHC recommendations~\cite{Butterworth:2015oua}. The $\alpha_s$ uncertainty is derived using the same PDF set but with two different $\alpha_s$ values. The uncertainties from the PDF and $\alpha_s$ variations are added in quadrature.
 
For the uncertainties due to parton shower and hadronization and due to the NLO matching, the nominal \PowPyttbar\ sample is compared with the \PowHwttbar\ sample.
To evaluate the effect of the amount of initial and final state radiation (ISR and FSR) on the measurement, the settings of the nominal \PowPyttbar~sample are varied.
The uncertainty due to ISR is estimated by simultaneously changing \mur\ and \muf\ in the ME and $\mu_{\mathrm{R}}^{\mathrm{ISR}}$ in the PS, while the uncertainty due to FSR is estimated by changing $\mu_{\mathrm{R}}^{\mathrm{FSR}}$ in the PS.
 
Certain regions in the calibration measurement contain a sizable contribution of up to 13\% of \ttbar events with additional heavy flavor quarks.
Various measurements of this process have shown that the simulation consistently underestimates the contribution~\cite{HIGG-2020-23}, therefore an additional
one-sided uncertainty of 28\% is introduced to account for this discrepancy.
 
Single-top-quark processes contribute sub-dominantly in some parts of the phase space and are governed by an additional uncertainty stemming from
different ways to treat the interference between \ttbar and $tW$ production. The nominal $tW$ samples use the diagram removal scheme, while the alternative samples use the diagram subtraction scheme~\cite{Frixione:2008yi,Re:2010bp}.


\subsection{Calibration results}


The nine calibration regions described in Section~\ref{sec:measurement} are included in a simultaneous maximum-likelihood fit using five unconstrained parameters (two \ttbar\ normalization parameters and the three POIs: \SFeff, \SFfakelow, \SFfakehigh). Systematic uncertainties, as described in Section~\ref{sec:systematics}, are treated in the fit as $\mathcal{O}(100)$ nuisance parameters with Gaussian prior. The width of the Gaussian, $\sigma$, is equal to the systematic uncertainty corresponding to the nuisance parameter. Uncertainties are inclusive in $B$-hadron \pt. All common sources of uncertainties are treated as correlated across eight of the nine measurement regions but uncorrelated with respect to the `0\ssv\ 2b, high-\met' region used to constrain the $\mu_{\ttbar}^{\text{high}-\met}$ normalization parameter. This treatment guarantees that no constraint is imposed on the POIs from the high-\met normalization region.
The \pileup\ uncertainties are the only exception to this treatment. These uncertainties are also treated as uncorrelated between regions with zero or at least one SSV and they are also uncorrelated between high and low $\mu$ regions.
 
\begin{figure}
\centering
\begin{subfigure}{.75\textwidth}
\includegraphics[width=.9\textwidth]{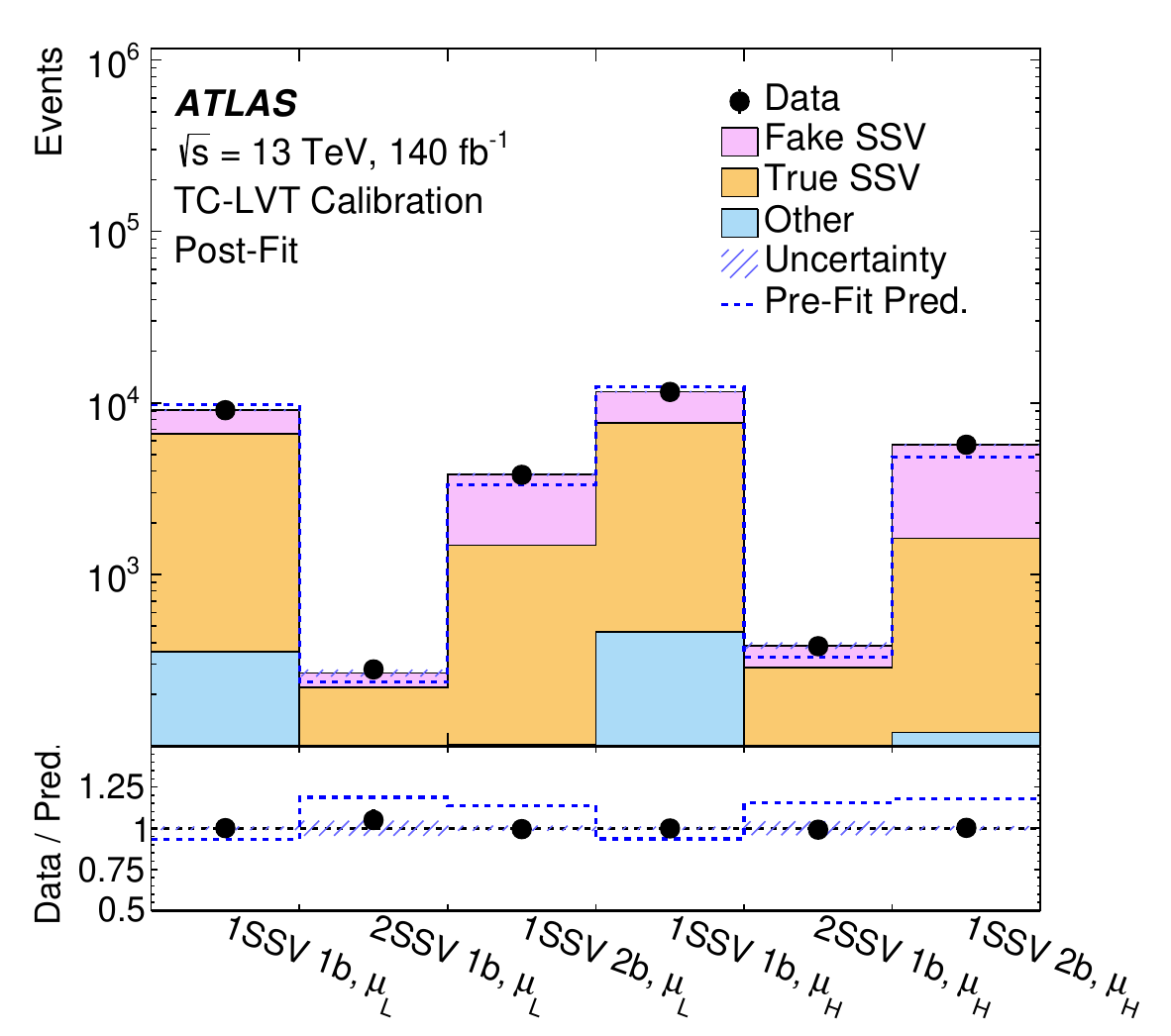}
\caption{\label{fig:postFit_1}}
\end{subfigure}
\begin{subfigure}{.75\textwidth}
\includegraphics[width=.9\textwidth]{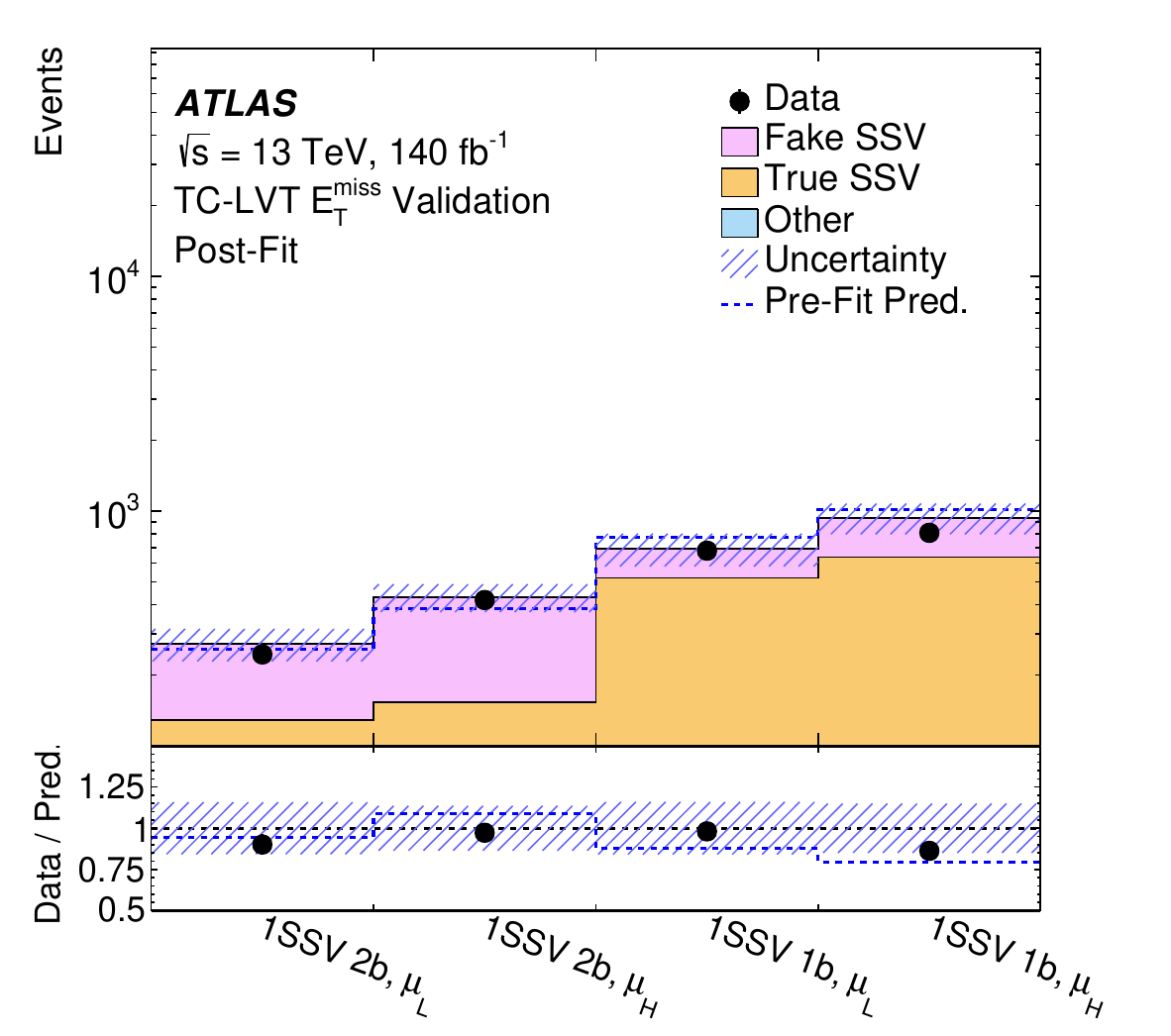}
\caption{\label{fig:postFit_2}}
\end{subfigure}
\caption{Summary of the fit results in \subref{fig:postFit_1} calibration regions and \subref{fig:postFit_2} validation regions. The prediction is separated into top quark events (i.e. \ttbar and single top quark processes) containing only true \ssv s (true) and events containing at least one fake \ssv ~(fakes). The `Other' category contains contributions from non-top quark processes. The bottom panel shows the ratio of the data and the pre-fit (dashed) and post-fit (points) prediction. The hashed band shows the full uncertainty of the prediction. }
\label{fig:regionsummary}
\end{figure}

After the fit, a good agreement between the observed data and the prediction is obtained. Post-fit yields and their uncertainties are summarized for the calibration regions in Table~\ref{tab:yieldresult}, the validation regions in Table~\ref{tab:yieldresult_validation} and shown in Figure~\ref{fig:regionsummary}. Both calibration regions and  $\text{high}$-$\met$ validation regions are included in the latter. The POIs of the fit, summarized in Table~\ref{tab:fitresults}, are extracted in the calibration regions and applied in the validation regions, which also show good post-fit agreement between the data and the prediction.
 
\begin{figure}
\centering
\begin{subfigure}{.31\textwidth}
\includegraphics[width=\textwidth]{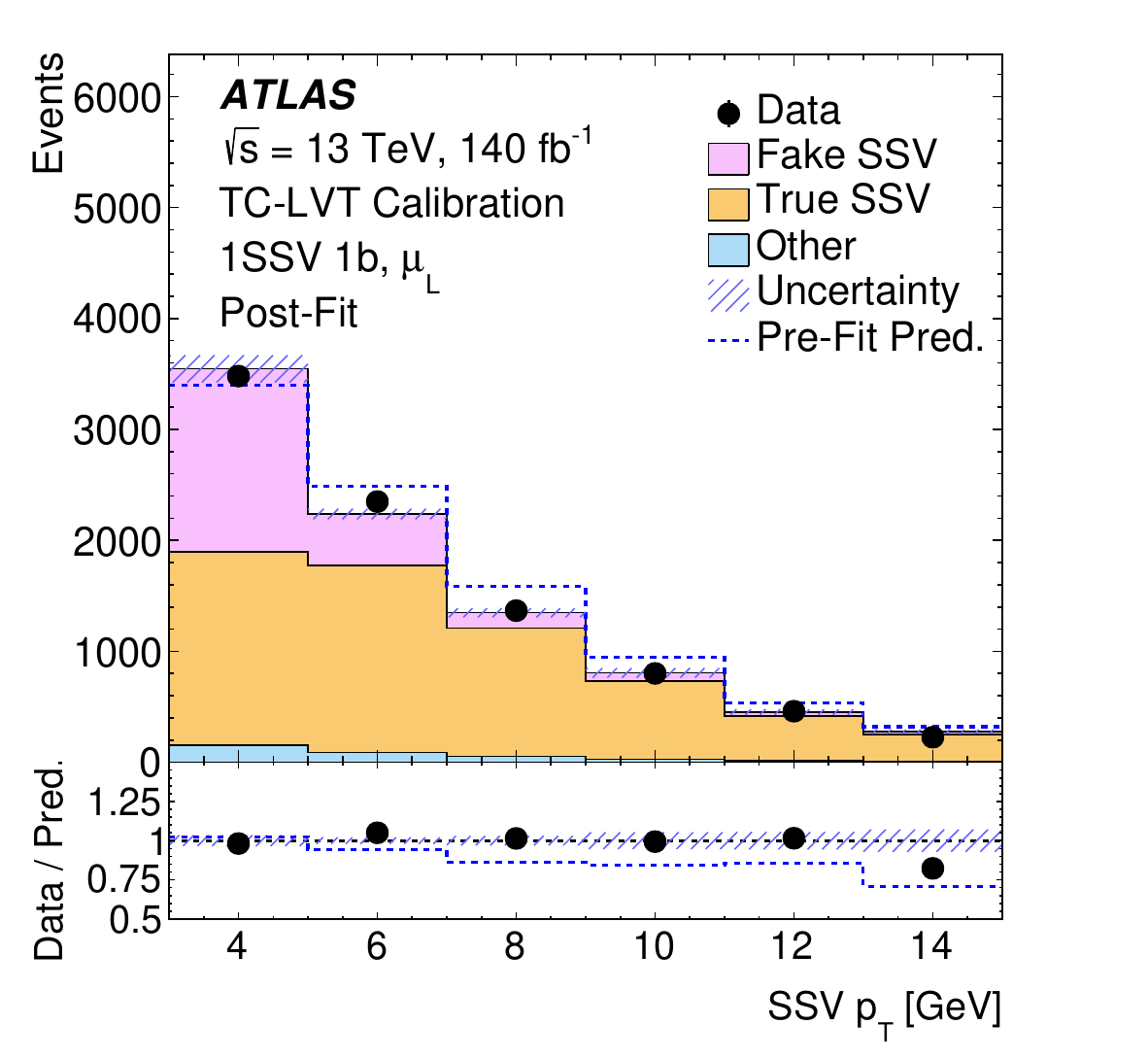}
\caption{\label{fig:postFit_pT_1}}
\end{subfigure}
\begin{subfigure}{.31\textwidth}
\includegraphics[width=\textwidth]{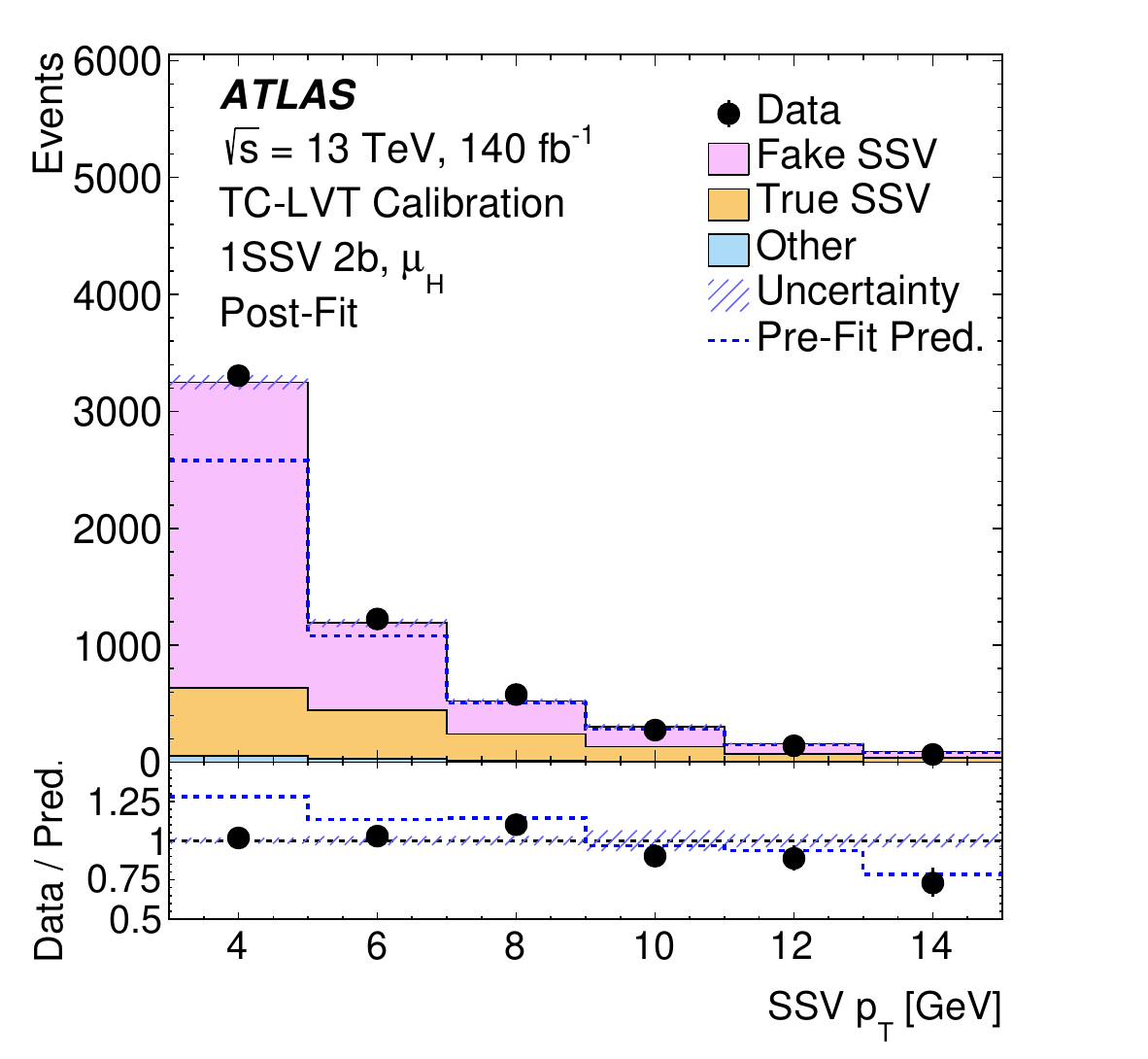}
\caption{ \label{fig:postFit_pT_2}}
\end{subfigure}
\begin{subfigure}{.31\textwidth}
\includegraphics[width=\textwidth]{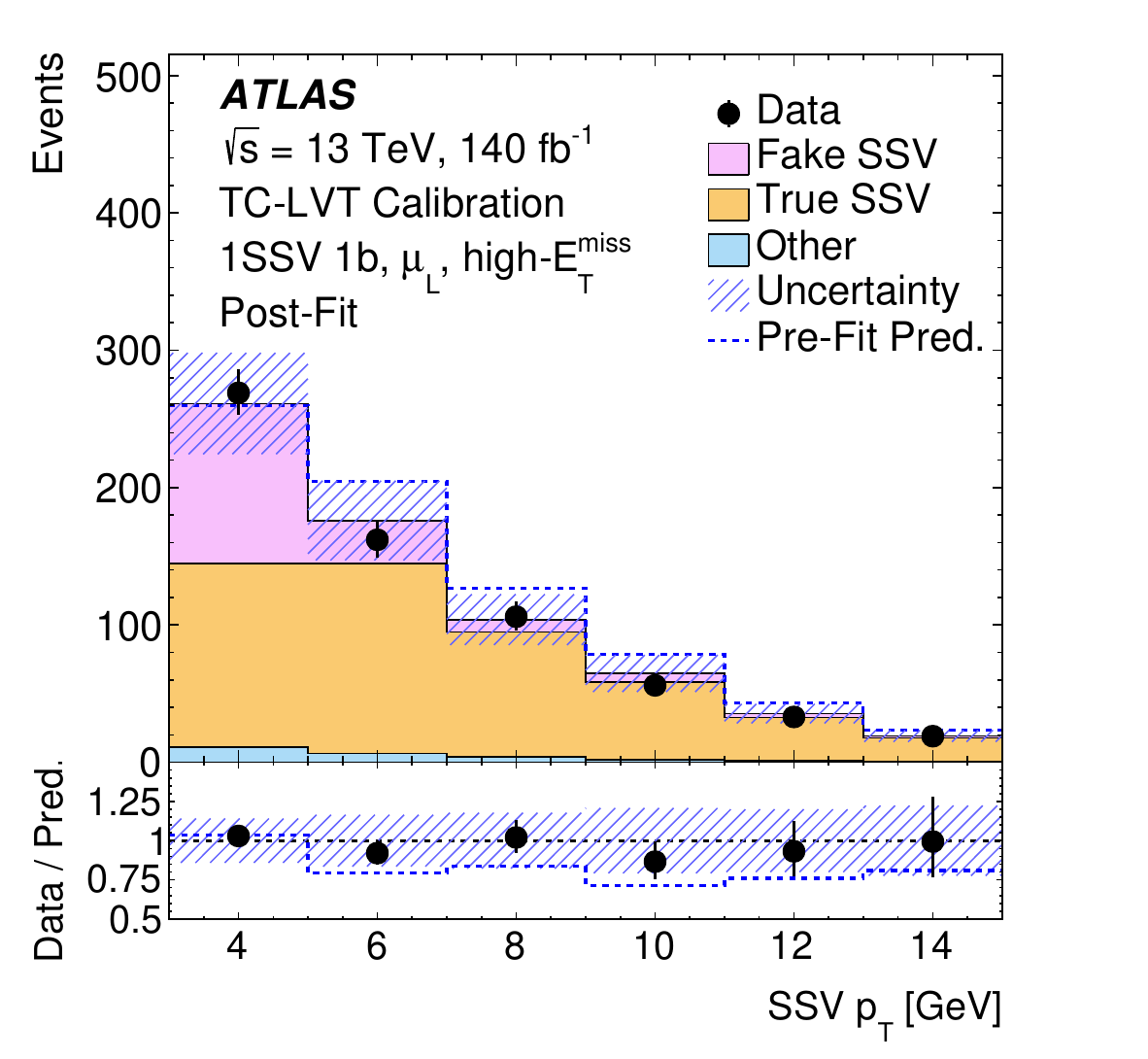}
\caption{ \label{fig:postFit_pT_3}}
\end{subfigure}
 
\caption{Distribution of the SSV \pt in \subref{fig:postFit_pT_1}, \subref{fig:postFit_pT_2} two calibration regions and \subref{fig:postFit_pT_3} one validation region. The prediction, which is shown post-fit, is separated into top quark events (i.e. \ttbar and single top quark processes) containing only true \ssv s (true) and events containing at least one fake \ssv ~(fakes). The `Other' category contains contributions from non-top quark processes. Hashed bands indicate the full uncertainty of the prediction. The last bin contains also the events in the overflow.}
\label{fig:kinematicMain}
\end{figure}

Selected post-fit distributions of the \ssv\ transverse momentum are shown in Figure~\ref{fig:kinematicMain} for a subset of the calibration and validation regions. This calibration targets events with \ssv\ transverse momentum smaller than $20\;\GeV$. Only a very small fraction of events considered in the calibration contain an SSV with a transverse momentum  above $15\; \GeV$. The overall good agreement between the observed data and the prediction as a function of $\pt$ indicates that the assumption of $\pt$-independent SFs holds and supports the applicability of this calibration to analyses characterized by SSV \pT spectra significantly different from that in the calibration sample.
The small discrepancy between the experimental data and the  simulation in the SSV \pt bin $13-15\; \GeV$, indicates a possible \pT-dependence of the calibration
for values of SSV \pt\ beyond $15\;\GeV$, which are provided as validity limit for the applicability of this calibration. Further investigations would be
needed in order to extend the calibration beyond this SSV \pt range.

 
\begin{table}
\centering
\caption{Post-fit yields, including systematic uncertainties, in all calibration regions. Due to correlations, systematic uncertainties on the individual samples do not necessarily add up in quadrature to the total uncertainty. \ttbar events are assigned to the `\ttbar fakes' category if the event contains at least one fake \ssv. }
\label{tab:yieldresult}
\begin{tiny}
\begin{tabular}{l|c|c|c|c|c|c|c|c}
\toprule
&  0\ssv\ 2b, $\mu_{\hbox{L}}$ &  0\ssv\ 2b, $\mu_{\hbox{H}}$ &  1\ssv\ 1b, $\mu_{\hbox{L}}$ &  2\ssv\ 1b, $\mu_{\hbox{L}}$ &  1\ssv\ 2b, $\mu_{\hbox{L}}$ &  1\ssv\ 1b, $\mu_{\hbox{H}}$ &  2\ssv\ 1b, $\mu_{\hbox{H}}$ &  1\ssv\ 2b, $\mu_{\hbox{H}}$\\\midrule
\ttbar true & 194940 $\pm$ 20150 & 225310 $\pm$ 21660 & 5760 $\pm$ 610 & 62 $\pm$ 10 & 1316 $\pm$ 164 & 6530 $\pm$ 640 & 63 $\pm$ 11 & 1447 $\pm$ 177\\\midrule
\ttbar fakes & -  & - & 2350 $\pm$ 387 & 177 $\pm$ 32 & 2348 $\pm$ 323 & 3697 $\pm$ 485 & 277 $\pm$ 38 & 4012 $\pm$ 423\\\midrule
Single top & 6814 $\pm$ 840 & 8161 $\pm$ 1006 & 781 $\pm$ 100 & 22 $\pm$ 5 & 132 $\pm$ 18 & 1102 $\pm$ 140 & 36 $\pm$ 7 & 207 $\pm$ 29\\\midrule
Others & 960 $\pm$ 190 & 1130 $\pm$ 230 & 180 $\pm$ 40 & 5 $\pm$ 1 & 30 $\pm$ 6 & 270 $\pm$ 60 & 7 $\pm$ 2 & 40 $\pm$ 8\\\midrule
Total predicted & 202400 $\pm$ 900 & 234700 $\pm$ 900 & 9060 $\pm$ 100 & 266 $\pm$ 12 & 3830 $\pm$ 60 & 11600 $\pm$ 110 & 384 $\pm$ 17 & 5690 $\pm$ 70\\\midrule
Data & 202359  & 234733  & 9068  & 279  & 3806  & 11589  & 381  & 5703 \\\midrule
 
\bottomrule\end{tabular}
\end{tiny}
 
\end{table}
 
\begin{table}
\centering
 
\caption{Post-fit yields, including systematic uncertainties, in all high-\met validation regions. The `$0\ssv\ 2b$' region is used to constrain the $\ttbar$ normalization in all validation regions.  Due to correlations, systematic uncertainties on the individual samples do not necessarily add up in quadrature to the total uncertainty. \ttbar events are assigned to the `\ttbar fakes' category if the event contains at least one fake \ssv.}
\label{tab:yieldresult_validation}
\begin{small}
\begin{tabular}{l|c|c|c|c|c}
\toprule
&  0\ssv\ 2b  &  1\ssv\ 2b, $\mu_{\hbox{L}}$  &  1\ssv\ 2b, $\mu_{\hbox{H}}$ &  1\ssv\ 1b, $\mu_{\hbox{L}}$  &  1\ssv\ 1b, $\mu_{\hbox{H}}$ \\
& high-\met & high-\met & high-\met & high-\met & high-\met \\\midrule
\ttbar true & 23730 $\pm$ 3010 & 112 $\pm$ 29 & 133 $\pm$ 36 & 471 $\pm$ 108 & 569 $\pm$ 125\\\midrule
\ttbar fakes & - & 141 $\pm$ 34 & 263 $\pm$ 55 & 164 $\pm$ 37 & 282 $\pm$ 57\\\midrule
Single top & 1618 $\pm$ 182 & 14 $\pm$ 3 & 27 $\pm$ 6 & 49 $\pm$ 9 & 73 $\pm$ 12\\\midrule
Others & 270 $\pm$ 50 & 4 $\pm$ 1 & 7 $\pm$ 1 & 10 $\pm$ 2 & 14 $\pm$ 3\\\midrule
Total predicted & 25600 $\pm$ 160 & 270 $\pm$ 40 & 430 $\pm$ 60 & 690 $\pm$ 110 & 940 $\pm$ 140\\\midrule
Data & 25601  & 245  & 418  & 678  & 808 \\\midrule
 
\bottomrule\end{tabular}
\end{small}
\end{table}


\newpage
 
\begin{figure}
\centering
\begin{subfigure}{.45\textwidth}
\includegraphics[width=\textwidth]{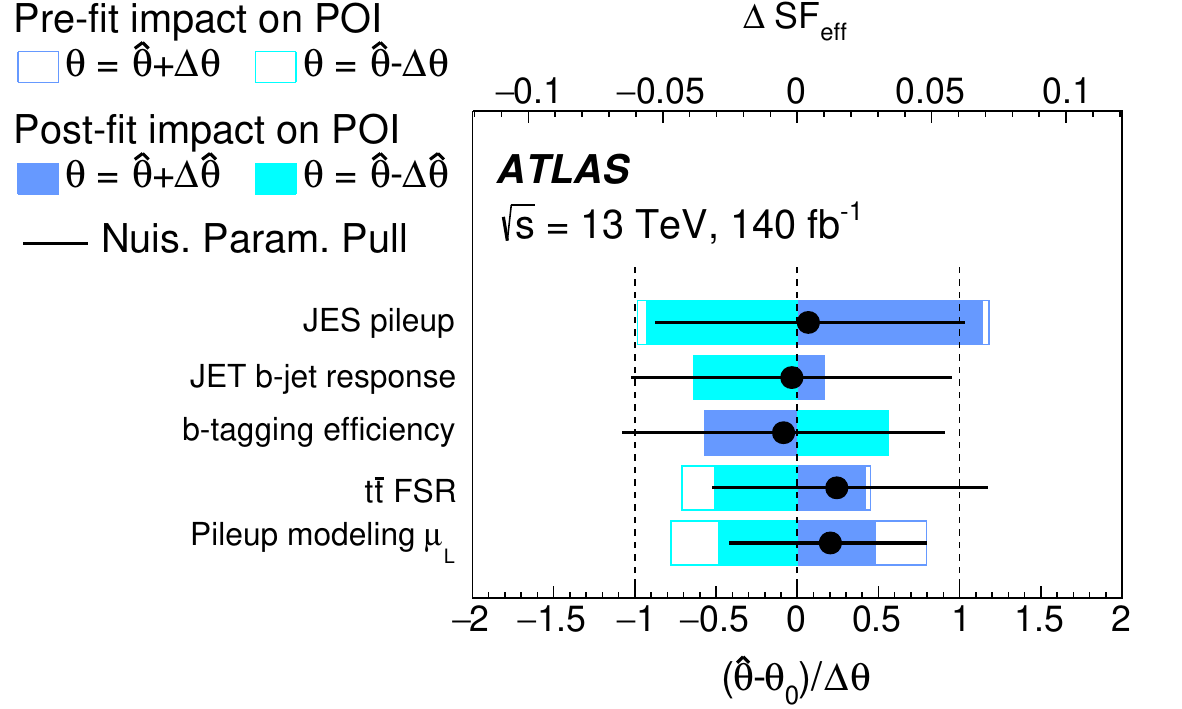}
\caption{}
\end{subfigure}
\begin{subfigure}{.45\textwidth}
\includegraphics[width=\textwidth]{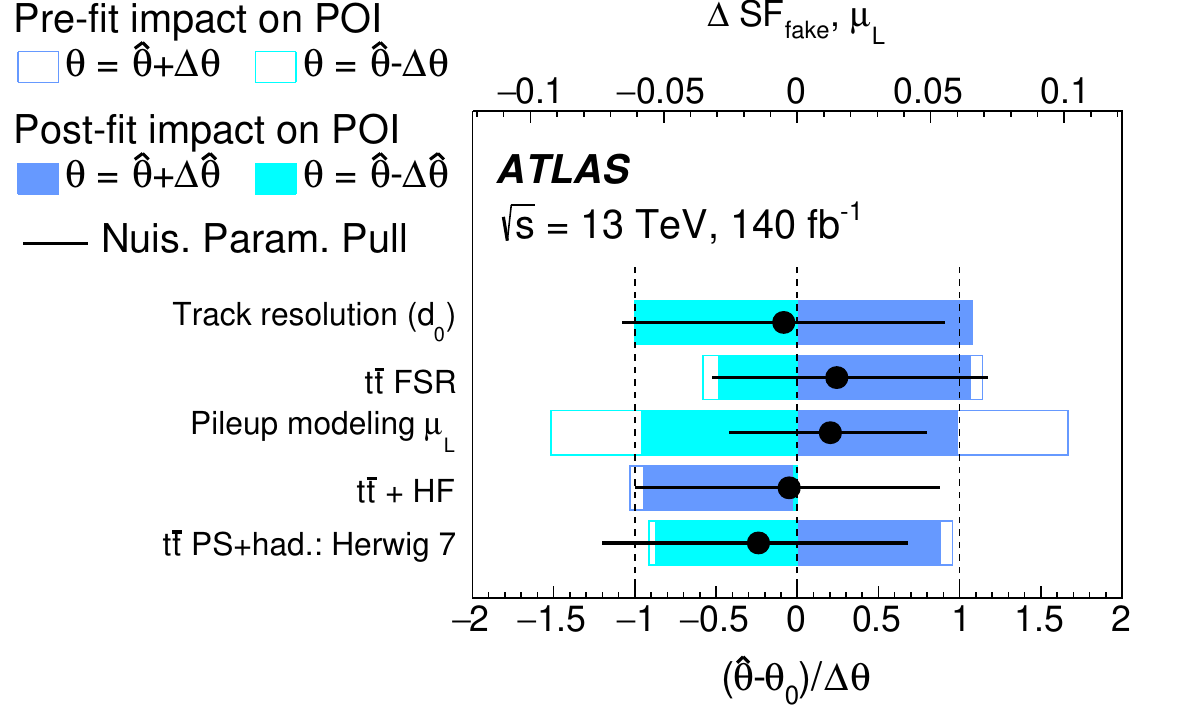}
\caption{}
\end{subfigure}
\begin{subfigure}{.45\textwidth}
\includegraphics[width=\textwidth]{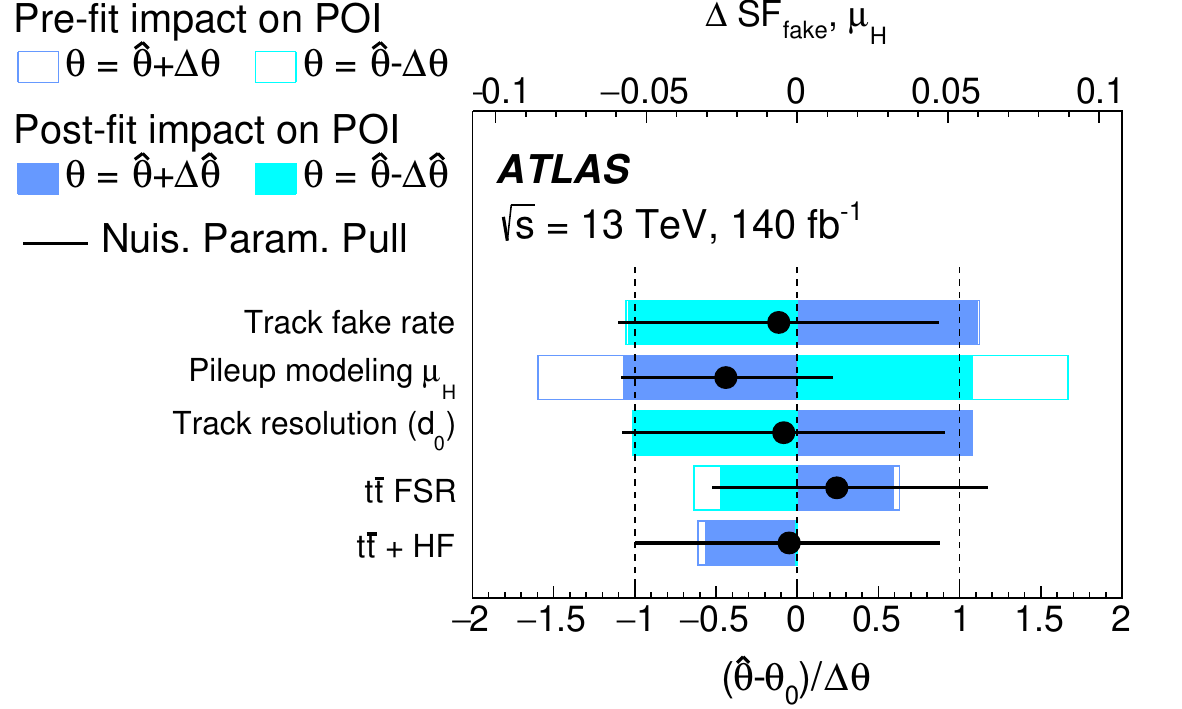}
\caption{}
\end{subfigure}
\caption{The five nuisance parameters with the largest impact on the three calibration scale factors (a) \SFeff, (b) \SFfakelow, and (c) \SFfakehigh.}
\label{fig:impact}
\end{figure}

\begin{table}
\centering
\caption{Fit results, with uncertainties, for the five unconstrained parameters. }
\label{tab:fitresults}
\begin{tabular}{|c|c|}
\toprule
Fitted parameter & value \\\midrule
\SFeff & $0.86$ $\pm$ $0.10$\\
\SFfakelow & $1.63$ $\pm$ $0.15$\\
\SFfakehigh & $1.58$ $\pm$ $0.13$\\\midrule
$\mu_{\ttbar}$ & $0.96$ $\pm$ $0.14$ \\
$\mu_{\ttbar}^{\text{high-}\met}$ & $0.91$ $\pm$ $0.19$\\
\bottomrule
\end{tabular}
 
\end{table}

\begin{table}
\centering
\caption{Summary of the impact of the dominant systematic uncertainties in this calibration for the three calibration scale factors. All systematic uncertainties are fully correlated bin-by-bin, except special cases as explained in the text. In the case of correlated systematic uncertainties, the relative sign of the uncertainty for each measurement is taken into account, even if not shown here.}
\label{tab:uncertainties}
\begin{tabular}{|l|c|c|c|}
\toprule
Uncertainty impact & \SFeff[\%] & \SFfakelow[\%] & \SFfakehigh[\%] \\
\midrule
Jet energy scale & 7 & 3 & 2 \\
Flavor tagging  & 3 & 3 & 2.5 \\
\ttbar modeling  & 3 & 6 & 3 \\
\ttbar heavy flavor modeling & 1.5 & 6 & 3 \\
Pileup & 3 & 6 & 6 \\
Track resolution & $<1$ & 6.5 & 6\\
Track fake rate & $<1$ &  $<1$ & 6 \\
\bottomrule
\end{tabular}
 
\end{table}
 
The efficiency scale factor is found to be smaller than one by about 15\%, while the two fake rate scale factors are found to be greater than one by about 60\% (Table~\ref{tab:fitresults}). The underlying causes of these effects are currently not fully understood, albeit these results are qualitatively consistent with those measured for \btagging\ in jets. The fitted values for the \ttbar\ normalization are consistent with factors obtained in previous measurements by the ATLAS Collaboration~\cite{SUSY-2018-08} targeting a similar phase space.
The nuisance parameters are only mildly constrained, with the exception of the parameters associated with the modeling of \pileup.
The fit constrains the \pileup uncertainties to about 50\% of its original value. This is expected because the \pileup uncertainties aim to cover the residual differences of data and simulation over an inclusive phase space, whereas the analysis bins are split into two separate $\mu$ regimes.
 
The impact of the five leading systematic uncertainties on the three POIs are presented in Figure~\ref{fig:impact}. A summary of the impact of all leading systematic uncertainties, grouped by uncertainty sources, is given in Table~\ref{tab:uncertainties}. Impacts are calculated by estimating, for each nuisance parameter,  the difference between the fitted POI in the nominal maximum-likelihood fit and a second fit performed by fixing the nuisance parameter to its nominal maximum-likelihood estimator plus one $\sigma$ of its Gaussian prior ($\hat{\theta}+1\sigma$).
Uncertainties related to the \pileup modeling and the modeling of extra-jet radiation in the \ttbar background are among the dominant sources of systematic uncertainties for all SFs.
Additionally, the uncertainties related to \bjet identification, jet energy scale, and jet flavor composition are important for \SFeff, while track-related uncertainties dominate for \SFfakelow\ and \SFfakehigh. Statistical uncertainties are below 1\% in this measurement.


\FloatBarrier
 
\section{Conclusion}
\label{sec:conclusion}
 
The reconstruction and identification of jets that originate from the hadronization process of $b$-quarks is an important component in
measurements and searches carried out by the ATLAS experiment. Various tagging algorithms and techniques are in place, covering
a large \bjet \pT range. A dedicated calibration of the TC-LVT algorithm, which focuses on the identification of
$B$-hadrons with such low \pt that they do not give rise to a reconstructed jet (\pT $< 20\;\GeV$), is presented for the first time in this paper.
The result of the calibration measurement is a set of scale factors that can be used to correct the performance of the TC-LVT algorithm in simulation to match that observed in data. These results improve and extend the previous calibration of this algorithm obtained in the context of bottom squark searches in the compressed mass regime.
 
The measurement is carried out using 140 fb$^{-1}$ of data from \pp collisions collected with the ATLAS detector during LHC Run~2 at a center-of-mass energy of 13~\TeV.
A data sample containing predominantly di-leptonic \ttbar events is selected, by requiring exactly one
muon and one electron of opposite electric charge. Various orthogonal regions are defined to facilitate the simultaneous measurement of efficiency and fake
rate scale factors. Simulated events are categorized based on the presence of true
or fake \ssv s, i.e. SSVs matched to a real $B$-hadron and SSVs mainly arising from random track crossings involving tracks from additional proton--proton interactions occurring in the same bunch crossing, reconstructed by the TC-LVT algorithm.

By separating events into regions with different
\bjet multiplicity, different number of SSVs and low and high \pileup, the different contributions of true and fake SSVs
in these regions can be exploited in a combined likelihood fit to extract the scale factors.
Auxiliary measurements are included as well, to reduce the dependence on any potential mis-modeling of the \ttbar process in the high missing transverse momentum regime.
Systematic uncertainties are included as nuisance parameters in the fit.
 
A single efficiency scale factor, \SFeff $=$ 0.86 $\pm$ 0.10, is determined. Two \pileup-dependent mistag rate scale factors,
\SFfakelow $=$ 1.63 $\pm$ 0.15 and \SFfakehigh $=$ $1.58$ $\pm$ 0.13, are measured. These scale factors are valid for SSV transverse momenta between
$3$ and $15\;\GeV$. The resulting efficiency in data for the TC-LVT  algorithm ranges from 1\%--12\% in the \pT range  3--15~\GeV. The TC-LVT algorithm and calibration described in this paper can significantly improve the sensitivity in various beyond-the-SM searches as they provide a novel and effective tool to reconstruct and identify events that contain low-momentum $b$-quarks.
 
\section*{Acknowledgements}


We thank CERN for the very successful operation of the LHC and its injectors, as well as the support staff at
CERN and at our institutions worldwide without whom ATLAS could not be operated efficiently.
 
The crucial computing support from all WLCG partners is acknowledged gratefully, in particular from CERN, the ATLAS Tier-1 facilities at TRIUMF/SFU (Canada), NDGF (Denmark, Norway, Sweden), CC-IN2P3 (France), KIT/GridKA (Germany), INFN-CNAF (Italy), NL-T1 (Netherlands), PIC (Spain), RAL (UK) and BNL (USA), the Tier-2 facilities worldwide and large non-WLCG resource providers. Major contributors of computing resources are listed in Ref.~\cite{ATL-SOFT-PUB-2023-001}.
 
We gratefully acknowledge the support of ANPCyT, Argentina; YerPhI, Armenia; ARC, Australia; BMWFW and FWF, Austria; ANAS, Azerbaijan; CNPq and FAPESP, Brazil; NSERC, NRC and CFI, Canada; CERN; ANID, Chile; CAS, MOST and NSFC, China; Minciencias, Colombia; MEYS CR, Czech Republic; DNRF and DNSRC, Denmark; IN2P3-CNRS and CEA-DRF/IRFU, France; SRNSFG, Georgia; BMBF, HGF and MPG, Germany; GSRI, Greece; RGC and Hong Kong SAR, China; ISF and Benoziyo Center, Israel; INFN, Italy; MEXT and JSPS, Japan; CNRST, Morocco; NWO, Netherlands; RCN, Norway; MNiSW, Poland; FCT, Portugal; MNE/IFA, Romania; MESTD, Serbia; MSSR, Slovakia; ARIS and MVZI, Slovenia; DSI/NRF, South Africa; MICINN, Spain; SRC and Wallenberg Foundation, Sweden; SERI, SNSF and Cantons of Bern and Geneva, Switzerland; NSTC, Taipei; TENMAK, T\"urkiye; STFC, United Kingdom; DOE and NSF, United States of America.
 
Individual groups and members have received support from BCKDF, CANARIE, CRC and DRAC, Canada; CERN-CZ, FORTE and PRIMUS, Czech Republic; COST, ERC, ERDF, Horizon 2020, ICSC-NextGenerationEU and Marie Sk{\l}odowska-Curie Actions, European Union; Investissements d'Avenir Labex, Investissements d'Avenir Idex and ANR, France; DFG and AvH Foundation, Germany; Herakleitos, Thales and Aristeia programmes co-financed by EU-ESF and the Greek NSRF, Greece; BSF-NSF and MINERVA, Israel; Norwegian Financial Mechanism 2014-2021, Norway; NCN and NAWA, Poland; La Caixa Banking Foundation, CERCA Programme Generalitat de Catalunya and PROMETEO and GenT Programmes Generalitat Valenciana, Spain; G\"{o}ran Gustafssons Stiftelse, Sweden; The Royal Society and Leverhulme Trust, United Kingdom.
 
In addition, individual members wish to acknowledge support from CERN: European Organization for Nuclear Research (CERN PJAS); Chile: Agencia Nacional de Investigaci\'on y Desarrollo (FONDECYT 1190886, FONDECYT 1230812, FONDECYT 1230987); China: Chinese Ministry of Science and Technology (MOST-2023YFA1605700), National Natural Science Foundation of China (NSFC - 12175119, NSFC 12275265, NSFC-12075060); Czech Republic: Czech Science Foundation (GACR - 24-11373S), Ministry of Education Youth and Sports (FORTE CZ.02.01.01/00/22\_008/0004632), PRIMUS Research Programme (PRIMUS/21/SCI/017); EU: H2020 European Research Council (ERC - 101002463); European Union: European Research Council (ERC - 948254, ERC 101089007), Horizon 2020 Framework Programme (MUCCA - CHIST-ERA-19-XAI-00), European Union, Future Artificial Intelligence Research (FAIR-NextGenerationEU PE00000013), Italian Center for High Performance Computing, Big Data and Quantum Computing (ICSC, NextGenerationEU); France: Agence Nationale de la Recherche (ANR-20-CE31-0013, ANR-21-CE31-0013, ANR-21-CE31-0022, ANR-22-EDIR-0002), Investissements d'Avenir Labex (ANR-11-LABX-0012); Germany: Baden-Württemberg Stiftung (BW Stiftung-Postdoc Eliteprogramme), Deutsche Forschungsgemeinschaft (DFG - 469666862, DFG - CR 312/5-2); Italy: Istituto Nazionale di Fisica Nucleare (ICSC, NextGenerationEU), Ministero dell'Università e della Ricerca (PRIN - 20223N7F8K - PNRR M4.C2.1.1); Japan: Japan Society for the Promotion of Science (JSPS KAKENHI JP21H05085, JSPS KAKENHI JP22H01227, JSPS KAKENHI JP22H04944, JSPS KAKENHI JP22KK0227); Netherlands: Netherlands Organisation for Scientific Research (NWO Veni 2020 - VI.Veni.202.179); Norway: Research Council of Norway (RCN-314472); Poland: Polish National Agency for Academic Exchange (PPN/PPO/2020/1/00002/U/00001), Polish National Science Centre (NCN 2021/42/E/ST2/00350, NCN OPUS nr 2022/47/B/ST2/03059, NCN UMO-2019/34/E/ST2/00393, UMO-2020/37/B/ST2/01043, UMO-2021/40/C/ST2/00187, UMO-2022/47/O/ST2/00148, UMO-2023/49/B/ST2/04085); Slovenia: Slovenian Research Agency (ARIS grant J1-3010); Spain: Generalitat Valenciana (Artemisa, FEDER, IDIFEDER/2018/048), Ministry of Science and Innovation (MCIN \& NextGenEU PCI2022-135018-2, MICIN \& FEDER PID2021-125273NB, RYC2019-028510-I, RYC2020-030254-I, RYC2021-031273-I, RYC2022-038164-I), PROMETEO and GenT Programmes Generalitat Valenciana (CIDEGENT/2019/023, CIDEGENT/2019/027); Sweden: Swedish Research Council (Swedish Research Council 2023-04654, VR 2018-00482, VR 2022-03845, VR 2022-04683, VR 2023-03403, VR grant 2021-03651), Knut and Alice Wallenberg Foundation (KAW 2018.0157, KAW 2018.0458, KAW 2019.0447, KAW 2022.0358); Switzerland: Swiss National Science Foundation (SNSF - PCEFP2\_194658); United Kingdom: Leverhulme Trust (Leverhulme Trust RPG-2020-004), Royal Society (NIF-R1-231091); United States of America: U.S. Department of Energy (ECA DE-AC02-76SF00515), Neubauer Family Foundation.


\printbibliography
 
\clearpage
 
\begin{flushleft}
\hypersetup{urlcolor=black}
{\Large The ATLAS Collaboration}

\bigskip

\AtlasOrcid[0000-0002-6665-4934]{G.~Aad}$^\textrm{\scriptsize 104}$,
\AtlasOrcid[0000-0001-7616-1554]{E.~Aakvaag}$^\textrm{\scriptsize 17}$,
\AtlasOrcid[0000-0002-5888-2734]{B.~Abbott}$^\textrm{\scriptsize 123}$,
\AtlasOrcid[0000-0002-0287-5869]{S.~Abdelhameed}$^\textrm{\scriptsize 119a}$,
\AtlasOrcid[0000-0002-1002-1652]{K.~Abeling}$^\textrm{\scriptsize 56}$,
\AtlasOrcid[0000-0001-5763-2760]{N.J.~Abicht}$^\textrm{\scriptsize 50}$,
\AtlasOrcid[0000-0002-8496-9294]{S.H.~Abidi}$^\textrm{\scriptsize 30}$,
\AtlasOrcid[0009-0003-6578-220X]{M.~Aboelela}$^\textrm{\scriptsize 45}$,
\AtlasOrcid[0000-0002-9987-2292]{A.~Aboulhorma}$^\textrm{\scriptsize 36e}$,
\AtlasOrcid[0000-0001-5329-6640]{H.~Abramowicz}$^\textrm{\scriptsize 154}$
% ...
\bigskip
\\

% ...
$^{17}$Department for Physics and Technology, University of Bergen, Bergen; Norway.\\
% ...
$^{30}$Physics Department, Brookhaven National Laboratory, Upton NY; United States of America.\\
% ...
$^{36}$$^{(a)}$Facult\'e des Sciences Ain Chock, Universit\'e Hassan II de Casablanca;$^{(b)}$Facult\'{e} des Sciences, Universit\'{e} Ibn-Tofail, K\'{e}nitra;$^{(c)}$Facult\'e des Sciences Semlalia, Universit\'e Cadi Ayyad, LPHEA-Marrakech;$^{(d)}$LPMR, Facult\'e des Sciences, Universit\'e Mohamed Premier, Oujda;$^{(e)}$Facult\'e des sciences, Universit\'e Mohammed V, Rabat;$^{(f)}$Institute of Applied Physics, Mohammed VI Polytechnic University, Ben Guerir; Morocco.\\
% ...
$^{45}$Physics Department, Southern Methodist University, Dallas TX; United States of America.\\
% ...
$^{50}$Fakult\"{a}t Physik , Technische Universit{\"a}t Dortmund, Dortmund; Germany.\\
% ...
$^{56}$II. Physikalisches Institut, Georg-August-Universit\"{a}t G\"ottingen, G\"ottingen; Germany.\\
% ...
$^{104}$CPPM, Aix-Marseille Universit\'e, CNRS/IN2P3, Marseille; France.\\
% ...
$^{119}$$^{(a)}$New York University Abu Dhabi, Abu Dhabi;$^{(b)}$United Arab Emirates University, Al Ain; United Arab Emirates.\\
% ...
$^{123}$Homer L. Dodge Department of Physics and Astronomy, University of Oklahoma, Norman OK; United States of America.\\
% ...
$^{154}$Raymond and Beverly Sackler School of Physics and Astronomy, Tel Aviv University, Tel Aviv; Israel.\\
% ...
$^{a}$ Also Affiliated with an institute covered by a cooperation agreement with CERN.\\
$^{b}$ Also at An-Najah National University, Nablus; Palestine.\\
$^{c}$ Also at Borough of Manhattan Community College, City University of New York, New York NY; United States of America.\\
$^{d}$ Also at Center for Interdisciplinary Research and Innovation (CIRI-AUTH), Thessaloniki; Greece.\\
$^{e}$ Also at Centro Studi e Ricerche Enrico Fermi; Italy.\\
$^{f}$ Also at CERN, Geneva; Switzerland.\\
% ...

\end{flushleft}


\clearpage

\end{document}